# Egocentric Mixed-Methods SNA: Analyzing Interviews with Women and/or LGBTQ+ Ph.D. Physicists

Authors: Chase Hatcher, Lily Donis, Adrienne Traxler, Madison Swirtz, Camila Manni, Justin Gutzwa, Charles Henderson, Ramón Barthelemy

Abstract: Social network analysis (SNA) has been widely used in physics education research (PER) in recent years, but mostly in a limited range of the available modalities. This paper describes a unique approach to egocentric, mixed-methods SNA applied to qualitative network data obtained from 100 interviews with women and/or queer professional physicists. We focus on our methods for obtaining quantitative network data from these qualitative sources and present novel techniques for analysis of the networks. We also examine the ways in which egocentric and mixed-methods SNA techniques are aligned with critical methods and well-suited to the study of difference, non-normativity, and experiences of marginalization in physics spaces and communities. We explore the limitations and potential applications of these methods and situate this work in the larger context of our study of these interviews. This work bridges a methodological gap between SNA and qualitative work on identity in PER and begins to develop our understanding of the way gender and sexual minority physicists experience support.

## I. Introduction

Social network analysis (SNA) is a set of tools and techniques developed in the social sciences for studying groups of people based on their connections. SNA has been used broadly in social science disciplines, including in physics education research (PER), a discipline in which multiple scholars have written guides promoting SNA's use as a method.[1,2,3,4] While SNA's quantitative versatility and deployment of spatial description may make it especially attractive to physicists, there are additional ways to use these tools for new and unique results.

---

[1] Eric Brewe, "The Roles of Engagement: Network Analysis in Physics Education Research," *Network Analysis*, n.d., 17.
[2] Remy Dou and Justyna P. Zwolak, "Practitioner's Guide to Social Network Analysis: Examining Physics Anxiety in an Active-Learning Setting," *Physical Review Physics Education Research* 15, no. 2 (July 3, 2019): 020105, https://doi.org/10.1103/PhysRevPhysEducRes.15.020105.
[3] Jesper Bruun, "Networks as Integrated in Research Methodologies in PER," 2016, 11–17, https://www.per-central.org/items/detail.cfm?ID=14295.
[4] Adrienne L. Traxler et al., "Person-Centered and Qualitative Approaches to Network Analysis in Physics Education Research," *Physical Review Physics Education Research* 20, no. 2 (October 21, 2024): 020132, https://doi.org/10.1103/PhysRevPhysEducRes.20.020132.

As Traxler et al. (2024) suggest in an earlier paper from this project, we are only consistently using part of the available modes of inquiry that SNA affords.[5] The authors find that most SNA studies in PER take a whole network approach, meaning that researchers try to construct networks that represent whole groups of people (like the students in a physics class), rather than adopting an egocentric network approach by constructing their networks around individuals. They also find that researchers in PER tend to use SNA in quantitative contexts, often not including the qualitative power that SNA affords.

This article offers a unique methodological approach to SNA that proposes using both egocentric and mixed methods SNA (MMSNA, mixing qualitative and quantitative approaches) due to their flexibility and ability to deal with nuanced data that reflect the complexity of personal experiences. Because of this ability to attend to personal experiences, we also argue that these expansive approaches to SNA methodologies are well-aligned with research guided by critical theories and disciplines (e.g. gender studies, queer theory, ethnic studies).

The specific methods we discuss are from a project which examined the professional support networks of gender and sexual minority (GSM) physicists.[6] We conducted interviews with 100 women and/or LGBTQ+ PhD-holding physicists in which participants discussed their career trajectories and support networks that influenced their professional journeys. They also created a visual artifact representing the people and communities in their personal networks. The network analysis component is one of several qualitative and mixed methods analytical approaches our team is applying to this dataset. The study aims to better identify and understand the systems of support which successfully retained GSM physicists through a PhD so we might provide recommendations to institutions and professional societies to better support these physicists going forward.

This paper only explores and explains the methods we developed for this project. It does not interpret or compare the networks—that will be the subject of a forthcoming results paper. We begin by briefly introducing and discussing critical theory (Section II. A), which is the theoretical basis we use to approach socio-cultural questions around diversity, equity, and inclusion (DEI) in the larger project. Next, we provide background on SNA (Section II. B) and examine how it has been used in PER, concluding with an examination of egocentric and MMSNA and a discussion of why they are the most appropriate approach to critically-aligned SNA work. We also provide background on the larger study (Section II. C) before describing the methods we are using to construct and analyze our networks (Section III). We conclude with a discussion of novelty, limitations, and potential applications of our method and forthcoming work on this project (Section IV).

---

[5] Traxler et al.

[6] Justin A. Gutzwa et al., "How Women and Lesbian, Gay, Bisexual, Transgender, and Queer Physics Doctoral Students Navigate Graduate Education: The Roles of Professional Environments and Social Networks," *Physical Review Physics Education Research* 20, no. 2 (September 12, 2024): 020115, https://doi.org/10.1103/PhysRevPhysEducRes.20.020115.

# II. Background & Theory

## A. Critical Theory

Critical theory is a mode of inquiry and body of sociocultural theory that encompasses sub-research areas including but not limited to queer theory, feminist theory, and critical race theory. That is not to say that these modes of inquiry have not come into conflict or disagreement with one another, but that they are often cited together as examples of critical work. As Max Horkheimer, a very early critical theorist has said, critical theory is "dominated at every turn by a concern for reasonable conditions of life," (p. 199) especially the betterment of life conditions for oppressed peoples.[7] Critical theories provide a way to critique, analyze, and understand power systems and structures and call attention to the way those systems and structures control, subjugate, and exploit their constituents and surroundings.[8] They also call attention to the socially constructed nature of widely held social norms and fixtures like gender, race, and patriarchy, inviting us to consider whether such norms are natural and universal or socially constructed and context-dependent. In seeking to understand and report on complex socio-cultural phenomena like support for marginalized physicists, we choose to let critical theories inform our approach so that we might prioritize the needs of oppressed groups, interrogation of powerful systems, and emancipatory aims.

Although the methods presented in this paper are in some ways informed by critical theories, the design and goals of the larger project are informed by critical theories more directly. See an earlier publication from this project, Gutzwa et al. (2024), for a review of the way critical theories have led to a framework that complements the project's dataset overall, especially in our understanding of science identity.[9] Methodological trends in critical work have still shaped the methods presented here, and through the remainder of this section we explain those connections in an effort to show the alignment of these (not explicitly critical) methods with critical methods.

Critical theories share approaches to research that prioritize personal experience and voice, especially in understanding unique experiences of minoritization.[10] We see a strong motivation to uncover the theoretical underpinnings and historical context of powerful norms and social systems—often an answer to the question of "who benefits from it being this way?" These ideologies manifest in critical research methods in a few ways. First, there is a need to work with qualitative data, as these are the only data that could fully preserve the thoughts and feelings of study participants. Second, it is important for study participants to have as much agency as possible in the study process so that the study can actually serve to better their living conditions.[11] Finally, a critical research design should include an acknowledgement of the researcher's subjectivity and reflection on the way the researcher's identities and experiences could have influenced the research process. This can be accomplished through a description of positionality and personal stake in a project, clear reporting about the exchange of information and ideas between researcher and study participants, and inclusion of varied perspectives and experiences on the research team. Through the remainder of this section, we note aspects of our methods that are aligned with the facets of critical methods presented here.

---

[7] Max Horkheimer, *Critical Theory: Selected Essays* (New York: Continuum Pub. Corp, 1982).
[8] Robin Celikates and Jeffrey Flynn, "Critical Theory (Frankfurt School)," in *The Stanford Encyclopedia of Philosophy*, ed. Edward N. Zalta and Uri Nodelman, Winter 2023 (Metaphysics Research Lab, Stanford University, 2023), https://plato.stanford.edu/archives/win2023/entries/critical-theory/.
[9] Gutzwa et al., "How Women and Lesbian, Gay, Bisexual, Transgender, and Queer Physics Doctoral Students Navigate Graduate Education."
[10] Miguel Rodriguez, Ramón Barthelemy, and Melinda McCormick, "Critical Race and Feminist Standpoint Theories in Physics Education Research: A Historical Review and Potential Applications," *Physical Review Physics Education Research* 18, no. 1 (February 25, 2022): 013101, https://doi.org/10.1103/PhysRevPhysEducRes.18.013101.

## Author Positionality

Most members of the research team are either currently experiencing or have formerly experienced doctoral-level graduate programs in physics while holding minoritized identities; it is our own experiences navigating inequity and of seeing others experience intersecting modalities of domination and inequity that bring us to the present work. All authors but two (Gutzwa and Donis) completed or are currently completing doctoral degree programs in physics, applied mathematics, and/or physics education research. Gutzwa completed a doctoral degree in higher education and Donis is currently completing an undergraduate degree. As such, our professional and personal experiences in physics and higher education more generally have heavily shaped our interpretation of findings, as well as the ways that Gutzwa, Manni, and Swirtz engaged with participants during data collection. Our varying goals and interests align in dedication to improving equity in STEM and thus serve the ultimate goal of this project.

Beyond the general commonalities of our experiences, our individual experiences and identities also shaped our engagement with the findings shared in this manuscript. Chase (Hatcher, he/him) is a white PhD candidate in physics with a background in physics education research and an interest in teaching physics at the university level after completing his degree. He joined this project after data collection and the first round of qualitative coding were complete and has had no interaction with the study participants. He does not embody any identities commonly marginalized in physics and is motivated to undertake this work less by personal experiences with inequity and more by observations of inequity as experienced by others. He is interested in SNA as a means of studying communities of learners with a focus on relationships, emancipatory approaches to STEM education, and making physics communities more welcoming for diverse physicists. His work on this project represents a confluence of these interests as well as a means of serving his teaching goals by providing material for his dissertation. Lily (Donis, she/her) is a white, disabled, woman currently completing an interdisciplinary undergraduate degree focused on public health with plans to pursue a doctoral degree. While her career and educational goals do not include physics, she works in a physics education research group and was highly motivated by her studies to approach the quantitative analysis in this project with a multitude of perspectives. Justin (Gutzwa, they/them) identifies as a white, trans, nonbinary, queer, disabled scholar-practitioner, and currently serves as an assistant professor in a college of education at a large, public university in the United States. Their status as a higher education scholar was often discussed in interviews as a mode of disrupting the traditional expectations of identity avoidance that participants might have held entering a STEM education study, and their prior experiences navigating doctoral programs as a trans student and advocating for racial and gender-based equity reformed within the microcosms of the research teams they worked on as a graduate student created avenues for shared empathy with the experiences participants shared during interviews. Ramón (Barthelemy, he/him) is a queer Hispanic physics education researcher who has worked on issues of gender and LGBTQ+ identity in physics and astronomy for over a decade. He came to this work as both a physicist and member of the community being studied. Camila (Manni, she/her) has been read as white in her home country and since coming to the United States is easily recognized as Latina, experiencing part of the racism and exclusionary behavior typically directed towards Latina/o people. Her identity as a queer Woman of Color was perceived during the interviews with people with shared identities, which to a certain extent shaped the process of collecting and analyzing data. Madison (Swirtz, they/them) is a queer white graduate student in physics education research, and as such is currently navigating the struggles of being visibly queer as a graduate student in a physics department. Several participants viewed them as a young researcher to be nurtured, and as such spoke about their graduate experiences as a way to give advice and offer mentorship. Adrienne (Traxler, she/they) is a white physicist. Her career trajectory (physics to education to applied mathematics to physics to science education) has sometimes placed her in a peripheral position or required careful framing to find jobs in physics departments, which shapes her reading of participants' comments on career transitions. Lastly, Charles (Henderson, he/him) is a white male physics education researcher who does not identify in the communities discussed here, but has been an active mentor and ally to the co-authors conducting this work, and more broadly in the physics education research community.

[11] Patti Lather, "Research as Praxis," *Harvard Educational Review* 56, no. 3 (September 1, 1986): 257–78, https://doi.org/10.17763/haer.56.3.bj2h231877069482.

## Critical Work in PER

Critically-informed work is not new to PER—here we briefly review some key papers in critical PER to highlight their implications for methods and to serve as resources in connecting critical theories to PER. Traxler et al. (2016) proposes a more nuanced treatment of gender in PER and provides an introduction to the ideas of gender performativity, an introduction to intersectionality, and a review of the ways researchers in PER have studied gender.[12] They suggest transcending the "gender gap" framework so common in PER and doing qualitative work around gender. The first of these suggestions has informed the design of this project overall, and the second has led to a mixed-methods (mixing qualitative and quantitative) approach to this study, which we explain in more detail in the next subsection. Swirtz and Barthelemy (2022) examines queer methods in PER and suggests ways to further queer both quantitative and qualitative methods.[13] Some suggestions include allowing self-identification in surveys, avoiding normalization to majority populations, breaking boundaries between participants and researchers, and incorporating researcher and participant positionality in analysis, all of which have been taken up by the larger project and/or by this study. Rodriguez et al. (2022) discusses critical race theory and feminist standpoint theory in general and in the context of PER and the theories' implications for both qualitative and quantitative work in PER.[14] Among other suggestions, the authors emphasize the centrality of experiential knowledge of the oppressed and reporting results from participants' words and perspectives, both of which are key features of the methods presented here.

[12] Adrienne L. Traxler et al., "Enriching Gender in PER: A Binary Past and a Complex Future," *Physical Review Physics Education Research* 12, no. 2 (August 1, 2016): 020114, https://doi.org/10.1103/PhysRevPhysEducRes.12.020114.

There are more examples of critical work in PER and more researchers engaging with critical theorists, including work on Black physics students and race in physics by Hyater-Adams, Rosa, and Cochran,[15,16,17,18] work on gender in PER,[19,20,21,22] and work on LGBTQ+ physicists.[23,24,25,26,27,28] This is not meant to be an exhaustive review of such work and authors, just a review of studies offering commentary on critically-aligned methods that have direct relevance to this study. In the next section, we look at our methodological approach, SNA, more closely and discuss choices that better align this approach with critical methods as described here.

---

[13] Madison Swirtz and Ramón Barthelemy, "Queering Methodologies in Physics Education Research," in *Physics Education Research Conference 2022*, PER Conference (Grand Rapids, MI, 2022), 457–62.
[14] Rodriguez, Barthelemy, and McCormick, "Critical Race and Feminist Standpoint Theories in Physics Education Research."
[15] Simone Hyater-Adams et al., "Critical Look at Physics Identity: An Operationalized Framework for Examining Race and Physics Identity," *Physical Review Physics Education Research* 14, no. 1 (June 1, 2018): 010132, https://doi.org/10.1103/PhysRevPhysEducRes.14.010132.
[16] Katemari Rosa, "Educational Pathways of Black Women Physicists: Stories of Experiencing and Overcoming Obstacles in Life," *Physical Review Physics Education Research* 12, no. 2 (2016), https://doi.org/10.1103/PhysRevPhysEducRes.12.020113.
[17] Geraldine L. Cochran, Theodore Hodapp, and Erika E. Alexander Brown, "Identifying Barriers to Ethnic/Racial Minority Students' Participation in Graduate Physics," in *2017 Physics Education Research Conference Proceedings* (2017 Physics Education Research Conference, Cincinnati, OH: American Association of Physics Teachers, 2018), 92–95, https://doi.org/10.1119/perc.2017.pr.018.
[18] Katemari Rosa et al., "Resource Letter RP-1: Race and Physics," *American Journal of Physics* 89, no. 8 (August 1, 2021): 751–68, https://doi.org/10.1119/10.0005155.
[19] Ramón S. Barthelemy, "Physics Education Research: A Research Subfield of Physics with Gender Parity," *Physical Review Special Topics - Physics Education Research* 11, no. 2 (2015), https://doi.org/10.1103/PhysRevSTPER.11.020107.
[20] Ramón S. Barthelemy, Melinda McCormick, and Charles Henderson, "Gender Discrimination in Physics and Astronomy: Graduate Student Experiences of Sexism and Gender Microaggressions," *Physical Review Physics Education Research* 12, no. 2 (August 1, 2016): 020119, https://doi.org/10.1103/PhysRevPhysEducRes.12.020119.
[21] Ramón S. Barthelemy, Melinda McCormick, and Charles R. Henderson, "Understanding Women's Gendered Experiences in Physics and Astronomy through Microaggressions," 2014, 35–38, https://www.per-central.org/items/detail.cfm?ID=13442.
[22] Ramón S. Barthelemy et al., "Educational Supports and Career Goals of Five Women in a Graduate Astronomy Program," *Physical Review Physics Education Research* 16, no. 1 (April 21, 2020): 010119, https://doi.org/10.1103/PhysRevPhysEducRes.16.010119.

## B. Social Network Analysis

Social network analysis (SNA) methods characterize groups of people by analyzing the connections between the people, certain qualities of the people, and sometimes the qualities of the connections themselves. We can describe characteristics of individuals in networks by examining their relationships to the rest of the network, or we can describe characteristics of the entire network.[29] SNA incorporates elements of graph theory, which is a body of knowledge and algorithms concerning ways to arrange diagrams to convey certain information, particularly in representing networks visually.[30] Graphical representations of social networks, often referred to as sociograms or simply graphs, feature nodes (or people or actors) as solid shapes like dots, and edges (or connections or ties) as lines connecting them. By representing networks using different methods and arrangement algorithms, we can encode information and highlight different features of the networks, like tight-knit groups of people, particularly well-connected people, or particularly isolated people.[31]

SNA often involves the use of quantitative metrics that describe various real aspects of networks or individuals. Some of these are whole network measures or structural properties such as global density, which is the ratio of the total number of edges to the total number of possible edges in a network. Density may tell us how quickly a transferable entity like information could travel through an entire network (higher density meaning more potential for quick transfer).[32] Perhaps more commonly, SNA is used to quantify characteristics of individuals in the network, or node properties. Many node properties are measures of centrality, a family of measures that capture the prominence of individual nodes within a network based on their position in the network. It can signify connectedness, importance, influence, or other qualities about a person depending on the measure being used and the nature of the network. For example, degree is the total number of ties a node has, so we could think of it as a way of measuring the extent of an individual's immediate connectedness within a network. As another example, betweenness is the number of shortest paths between pairs of other nodes that pass through a given node, so it relates to the importance of being an intermediary between others in a network.[33] These examples illustrate how different ways of measuring centrality might capture different dimensions of prominence or importance in a network and are a sample of the wide variety of quantitative SNA metrics.

---

[23] Ramón S. Barthelemy, "LGBT+ Physicists Qualitative Experiences of Exclusionary Behavior and Harassment," *European Journal of Physics* 41, no. 6 (October 2020): 065703, https://doi.org/10.1088/1361-6404/abb56a.
[24] Ramón S. Barthelemy et al., "LGBT+ Physicists: Harassment, Persistence, and Uneven Support," *Physical Review Physics Education Research* 18, no. 1 (March 28, 2022): 010124, https://doi.org/10.1103/PhysRevPhysEducRes.18.010124.
[25] Ramón S. Barthelemy et al., "Research on Gender, Intersectionality, and LGBTQ+ Persons in Physics Education Research," in *The International Handbook of Physics Education Research: Special Topics*, ed. Mehmet Fatih Taşar and Paula R. L. Heron (AIP Publishing LLC, 2023), https://doi.org/10.1063/9780735425514.
[26] Ramón S. Barthelemy et al., "Workplace Climate for LGBT+ Physicists: A View from Students and Professional Physicists," *Physical Review Physics Education Research* 18, no. 1 (June 13, 2022): 010147, https://doi.org/10.1103/PhysRevPhysEducRes.18.010147.
[27] Timothy J. Atherton et al., "LGBT Climate in Physics: Building an Inclusive Community" (College Park, MD: American Physical Society, 2016).
[28] Xandria R. Quichocho, Erin M. Schipull, and Eleanor W. Close, "Understanding Physics Identity Development through the Identity Performances of Black, Indigenous, and Women of Color and LGBTQ+ Women in Physics," 2020, 412–17, https://www.per-central.org/items/detail.cfm?ID=15518.
[29] Stephen P. Borgatti et al., "Network Analysis in the Social Sciences," *Science* 323, no. 5916 (February 13, 2009): 892–95, https://doi.org/10.1126/science.1165821.
[30] Alexandra Marin and Barry Wellman, "Social Network Analysis: An Introduction," in *The SAGE Handbook of Social Network Analysis*, ed. John Scott and Peter J. Carrington (SAGE Publications, 2014), https://methods.sagepub.com/book/the-sage-handbook-of-social-network-analysis.
[31] Ulrik Brandes, Patrick Kenis, and Jörg Raab, "Explanation through Network Visualization," *Methodology* 2, no. 1 (January 2006): 16–23, https://doi.org/10.1027/1614-2241.2.1.16.

## SNA in PER

PER researchers have developed their own guides for using SNA and have started calling on the community to explore applications of this set of tools.[34,35,36] A common use case is networks of students based on their enrollment in the same class. In a 2017 study by Zwolak and collaborators, students that were more central in "meaningful interaction" networks developed in their modeling instruction physics class were more likely to continue in physics.[37] Brewe and coauthors found that, in a network of university physics students who helped each other with homework in a physics learning center, gender and ethnicity variables did not predict centrality, indicating that this learning center was perhaps less exclusive than other physics spaces.[38] Other studies have investigated student networks based on laboratory interaction,[39] collaboration on repeated group exams,[40] and long-standing friendships.[41] In addition to networks comprised of students, researchers in PER have also studied networks comprised of faculty and based on their discussions of teaching with each other[42] and types of interaction that teaching assistants have with students.[43]

[32] Robert A. Hanneman and Mark Riddle, "Concepts and Measures for Basic Network Analysis," in *The SAGE Handbook of Social Network Analysis*, ed. John Scott and Peter J. Carrington (London ; Thousand Oaks, Calif: SAGE, 2011).
[33] Peter J Carrington, John Scott, and Stanley Wasserman, "Models and Methods in Social: Network Analysis," 2005, 345.
[34] Brewe, "The Roles of Engagement: Network Analysis in Physics Education Research."
[35] Traxler et al., "Person-Centered and Qualitative Approaches to Network Analysis in Physics Education Research."
[36] Bruun, "Networks as Integrated in Research Methodologies in PER."
[37] Dou and Zwolak, "Practitioner's Guide to Social Network Analysis."
[38] Eric Brewe, Laird Kramer, and Vashti Sawtelle, "Investigating Student Communities with Network Analysis of Interactions in a Physics Learning Center," *Physical Review Special Topics - Physics Education Research* 8, no. 1 (January 12, 2012): 010101, https://doi.org/10.1103/PhysRevSTPER.8.010101.
[39] Meagan Sundstrom et al., "Examining the Effects of Lab Instruction and Gender Composition on Intergroup Interaction Networks in Introductory Physics Labs," *Physical Review Physics Education Research* 18, no. 1 (January 4, 2022): 010102, https://doi.org/10.1103/PhysRevPhysEducRes.18.010102.
[40] Steven F. Wolf et al., "Social Network Development in Classrooms," *Applied Network Science* 7, no. 1 (December 2022): 24, https://doi.org/10.1007/s41109-022-00465-z.
[41] Javier Pulgar et al., "Long-Term Collaboration with Strong Friendship Ties Improves Academic Performance in Remote and Hybrid Teaching Modalities in High School Physics," *Physical Review Physics Education Research* 18, no. 1 (June 13, 2022): 010146, https://doi.org/10.1103/PhysRevPhysEducRes.18.010146.

Researchers in PER are also employing different types of methods in their approach to SNA, which we could group into qualitative, quantitative, or mixed methods. For a detailed review of these methodological approaches to SNA in PER in recent years, see Traxler and collaborators' aforementioned review of SNA methodologies in PER, of which this paper is meant to be a continuation.[44] In it, the authors show that most of the recent SNA work in physics has been quantitative, which, while not a problem, does not afford the same analytical power for nuanced and complex social dynamics. For this reason, the authors have promoted the adoption of mixed methods SNA (MMSNA), which would employ some combination of quantitative and qualitative work. According to Crossley and Edwards, quantitative SNA is useful for characterizing and exploring patterns of connections, but not so much for understanding the mechanisms underlying those connections, which is where qualitative analysis can be most helpful.[45] Another way to frame the distinction is quantitative work giving an "outsider" perspective on the network, while qualitative work gives an "insider" perspective, since such work would usually be based on personal accounts from members of the network.

Traxler and collaborators' paper also examines the modes of network construction and representation that have been used in PER in recent years, focusing on sociocentric (whole network) and egocentric (person-centered) as the two main approaches. They note that the overwhelming majority of SNA studies in PER have used sociocentric approaches, which have tradeoffs in treatment of complexity and scope of study which would depend on study goals. We examine egocentric approaches more closely in the next subsection.

## Egocentric SNA

The most common way to perform SNA in PER is in a sociocentric (or whole network) study, which means the network represents the complete set of connections among a group of people. Most of the examples we have presented so far have been sociocentric studies—in education contexts, this might mean that the researchers use a survey to collect connection information from each individual in a classroom in order to represent the whole network. Sociocentric methods are useful in cases where the network may be clearly bounded (like a classroom or a school) and are ideal for answering questions that pertain to those whole groups, for example about clustering tendencies among students after assigning them new groups to work in or about centrality as it may be related to scores on a test.

---

[42] Kathleen Quardokus and Charles Henderson, "Promoting Instructional Change: Using Social Network Analysis to Understand the Informal Structure of Academic Departments," *Higher Education: The International Journal of Higher Education Research* 70, no. 3 (September 2015): 315–35, https://doi.org/10.1007/s10734-014-9831-0.

[43] Joe Olsen et al., "Characterizing Social Behavior Patterns in Teaching Assistant Interactions with Students," *Physical Review Physics Education Research* 19, no. 2 (September 13, 2023): 020129, https://doi.org/10.1103/PhysRevPhysEducRes.19.020129.

[44] Traxler et al., "Person-Centered and Qualitative Approaches to Network Analysis in Physics Education Research."

[45] Nick Crossley and Gemma Edwards, "Cases, Mechanisms and the Real: The Theory and Methodology of Mixed-Method Social Network Analysis," *Sociological Research Online*, May 31, 2016, https://doi.org/10.5153/sro.3920.

But sometimes we may be more interested in individuals and in the way their network affects them or the way they affect their network. In those cases, we may look to egocentric network analysis, which involves the study of social networks as they are defined by and around one person, called the ego. There are a few key differences in the affordances, limitations, and required perspectives of sociocentric and egocentric studies. Structurally, an egocentric network could be seen as a subset of a whole network, where one node's connections to others would comprise their ego network. Otherwise, the only structural difference between a whole network and an ego network is that an ego network necessarily has one node (ego) that is connected to all others (alters). A small example of an egocentric network is shown in Figure 1.

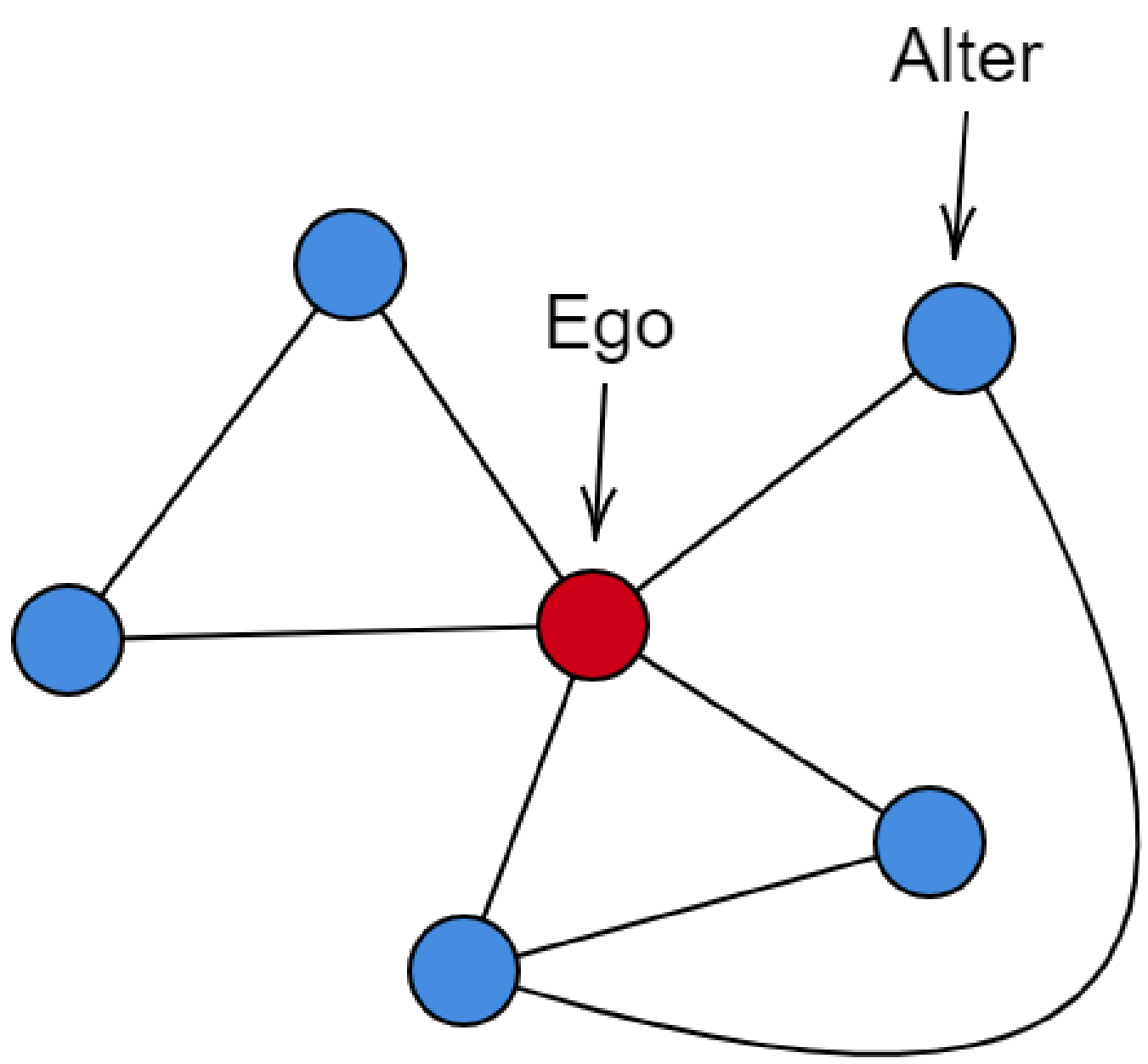

Figure 1: Example egocentric sociogram featuring connections between alters.

A key difference between sociocentric and egocentric approaches comes from the way the two types of networks are constructed. Whole networks are usually constructed from relational data gathered from each member of a group in which they describe their relationships with all other group members, whereas egocentric networks are constructed by the ego identifying all members of their network (alters), their relationships with them, and sometimes relationships among alters. They might also describe other characteristics of their alters, like age or race. In both cases, there is an assumption that an individual providing data can accurately describe their relationships with other people, but egocentric studies may have the added burden on the part of the ego of describing other people's relationships and demographics, which involves the same (and perhaps more) subjectivity, but not on the part of the subject. So, one limitation of egocentric studies is that all network information comes from one person, and they may not always be the best source for network characteristics that go beyond their immediate relationships.[46]

A benefit associated with using egocentric techniques is in the way it allows us to draw conclusions or make propositions from the data. Since the ego is only meant to be an example of some more general population, they actually serve as a sample of that population and afford the same generalizability as does any sample from a larger population.[47] On the other hand, the whole network is meant to represent the population itself, so there is no generalization to a larger population, only characterization of the specific population comprising the network, assuming the researchers can accurately capture the whole network. Neither generalizing from a sample or characterizing a population is necessarily better, but depending on the context, and especially when the population in question is too large to survey, the "sampling" perspective associated with egocentric methods is useful.

Though egocentric analysis is less common than sociocentric analysis, there are still a few relevant PER studies to review. A 2013 paper by Goertzen and co-authors describes their study of the effects of reforms implemented in introductory university physics, which they investigate by looking at three students' attitudes about learning physics, connections in the physics classroom, and relationships in the physics community.[48] The authors use egocentric network analysis after gathering information about students' ties and relationships, and link the students' expanded physics networks to successful reforms in the physics classes. It seems that the authors' choice to use egocentric methods has a lot to do with their desire to get an in-depth understanding of a few students' experiences in physics. Another study by Wu and collaborators examined differences in interactions between instructors and students in traditional and nontraditional (explorative) labs.[49] They used video and audio recording data of lab sessions to log interactions between the one instructor and multiple students in the class, thereby constructing egocentric networks with the instructor as ego. This is a unique example of egocentric data collection in which the subjectivity of the ego does not enter into the network construction—in other words, their data collection process more closely resembles that of a sociocentric study, except the one instructor is clearly the primary interactor with all students. So, in this example, the egocentric nature of the study may have more to do with the nature of the labs than with a methodological choice by the researchers.

---

[46] Brea L. Perry, Bernice A. Pescosolido, and Stephen P. Borgatti, *Egocentric Network Analysis: Foundations, Methods, and Models*, 1st ed. (Cambridge University Press, 2018), https://doi.org/10.1017/9781316443255.
[47] Perry, Pescosolido, and Borgatti.
[48] Renee Michelle Goertzen, Eric Brewe, and Laird Kramer, "Expanded Markers of Success in Introductory University Physics," *International Journal of Science Education* 35, no. 2 (2013): 262–88, https://doi.org/10.1080/09500693.2012.718099.
[49] David G. Wu et al., "Instructor Interactions in Traditional and Nontraditional Labs," *Physical Review Physics Education Research* 18, no. 1 (March 14, 2022): 010121, https://doi.org/10.1103/PhysRevPhysEducRes.18.010121.

## Critically-Aligned SNA

A goal of this project is to analyze the networks of study participants in a way that is attentive to how their intersecting dimensions of identity (some of which are marginalized in physics) have shaped their experiences. We turned to critical theories as a body of work that can inform our treatment of marginalized identities and explain here how those theories have shaped our approach to SNA. Specifically, we explain how egocentric and MMSNA, while not explicitly informed by critical theories, are aligned with some aspects of critical methodologies.

A common thread in critical work is the study of personal experiences with oppression in order to understand oppression, exemplified in standpoint theory and its variations as explained in greater detail (and in PER contexts) by Rodriguez et al.[50] It is easy to see why egocentric approaches to SNA, which literally center the individual in their own network, are aligned with this aspect of a critical methodology. By using egocentric approaches, we can ensure that the networks we construct are primarily if not completely articulated from the study participant's standpoint, ensuring that their perspective is privileged in our understanding of their experience. Another important tool of critical methodologies is storytelling, or the use of narrative data, which allows us to understand personal experiences in context and opens the door for counter-storytelling, or stories that challenge norms.[51] The only way that we could possibly include storytelling in an SNA study is by having a study design with large qualitative components, or MMSNA. By using (and in our case, prioritizing) qualitative descriptions of networks, we are able to include more story-based understandings of the networks. We are also able to potentially include mechanistic descriptions of our networks, or explanations of how and why certain aspects of the network are the way they are rather than just descriptions and characterizations of those aspects. This is also aligned with another common feature of critical studies, which is examining phenomena in context and with some understanding of how they came to be, not just how they are at one moment in time.

Critical methodologies also tend to feature a maximal amount of agency on the part of study participants. As we have discussed, this agency often comes in the form of dialogue (rather than monologue) during a data collection process like an interview, reciprocity in information sharing, and involvement in validation of results through checks and reviews. It also means that systemic and structural intuitions about the study matter should come directly from participants with little direction from the researcher—that is, whatever hypotheses or preconceived notions a researcher may have should not constrain the participants' voice. Using SNA presents a unique challenge here because it inherently involves the imposition of structure (the network) on an entity that may or may not be seen as having that structure independently of the study.[52] For example, if a researcher surveys a class of students on their connections and then constructs a network from their responses, is the network constructed by the students or the researcher, or a mix of both? We may see the hand of the researcher in at least some part of the network construction, which should give us pause, especially as critical theorists have focused so heavily on the construction of social structures as a way that power systems are maintained and perpetuated.

---

50 Rodriguez, Barthelemy, and McCormick, "Critical Race and Feminist Standpoint Theories in Physics Education Research."

51 "Critical Race Methodology : Counter-Storytelling as an Analytical Framework for Educational Research," in *Foundations of Critical Race Theory in Education* (Routledge, 2023), 159–74, https://doi.org/10.4324/b23210-18.

52 Louise Ryan, Jon Mulholland, and Agnes Agoston, "Talking Ties: Reflecting on Network Visualisation and Qualitative Interviewing," *Sociological Research Online* 19, no. 2 (May 1, 2014): 1–12, https://doi.org/10.5153/sro.3404.

If SNA generally has the drawback that it involves the researcher in the construction of structures which may or may not be otherwise present (according to the study participants/network constituents), it may be wise to work with SNA modalities that ensure the largest amount of agency on the part of the study participants in constructing the network. Egocentric approaches typically involve one person (the ego) describing their connections with and qualities of their alters, essentially building the network themselves. It may be the case that the structure the ego describes may not have existed for them until the researcher asked them to build it, meaning the researcher and their name generator had some role to play in the network's construction, but we could at least say that the burden of network construction lies with the study participant, which we might not be able to say for sociocentric analysis. Better, still, is if study participants are given the opportunity to visually construct their network as well, since the researchers using SNA will usually visualize the networks themselves. In fact, other studies on the use of visual tools in data collection have found that such techniques empower and give agency to users while also allowing for the articulation of unique perspectives.[53,54]

While SNA in general may not be designed to focus on things that critical theories focus on, like intersectionality, experiences with oppression and oppressive power systems, and challenges to widely held norms (the focus is instead on interpersonal connectivity), the best way that we can feature these topics in an SNA study is through qualitative data, so we again see a reason for using MMSNA in a critically-aligned SNA study design. Data on topics of power and oppression would more likely come from less SNA-oriented study design features (i.e. interview questions that probe these areas) and an eye for such discussions when analyzing the dataset, but the inclusion of qualitative data collection in the study design would be the bare minimum for allowing such topics to enter the analysis. The further inclusion of egocentric methods and visual tools for data collection, while again not explicitly derived from or informed by critical methods, is aligned with the person-centering techniques in critical methods and serves to make the combination of egocentric and MMSNA perhaps most appropriate of all SNA modalities for the study of normativity, difference, and minoritization in physics.

---

[53] Sofia Gameiro et al., "DrawingOut - An Innovative Drawing Workshop Method to Support the Generation and Dissemination of Research Findings," *PloS One* 13, no. 9 (2018): e0203197, https://doi.org/10.1371/journal.pone.0203197.

[54] Dawn Mannay, "Making the Familiar Strange: Can Visual Research Methods Render the Familiar Setting More Perceptible?," *Qualitative Research* 10, no. 1 (February 1, 2010): 91–111, https://doi.org/10.1177/1468794109348684.

## C. This Project

The larger project, of which this study is a part, is motivated by a noted gap in the PER community's understanding of what support looks like for GSM physicists—much of the research on the experiences of GSM physicists has been focused on their experiences of marginalization and discrimination, and the only study we have found that looks at experiences of support (McCormick et al. 2014) is focused on astronomy graduate students, not professional physicists.[55] Also, despite there being plenty of critical work in PER and SNA work in PER, we have not found any critical SNA work in PER. This project addresses the content gap on success and support of professional GSM physicists and the methodological gap between critical and SNA work in PER.

Our approach was to conduct and analyze interviews with women and/or LGBTQ+ PhD-holding physicists in different career sectors (government, industry, and academia). The interviews focused on career trajectories and experiences of support, and since we were also interested in our participants' professional networks, the interviews also served as egocentric network name generators. Much of the analysis our team is doing on this dataset is qualitative and based on thematic coding of the interview transcripts, but the present study takes an SNA approach to the dataset. By using critically-aligned SNA methods applied to qualitative egocentric network and interview data, we develop a comprehensive understanding of the way professional support networks function and differ among physicists in different career sectors and with different dimensions of personal identity. Our hope is that, with a better understanding of our study participants' support networks, we can make recommendations to institutions and professional societies on how to better support similar physicists going forward.

# III. Methods

In this section, we review our methods, starting with our interview process and focusing on the parts of that process oriented towards network construction. Next, we describe our thematic coding process, again focusing on aspects most relevant to network construction and analysis. Then, we describe our data processing from thematic codes to .csv files more usable in coding environments, and we describe the computer coding that leads to visual and statistical analysis of our dataset.

## A. Interviews

Our dataset consists of transcripts of recorded interviews conducted with 100 women and/or LGBTQ+ PhD-holding physicists. Our interviews served as both a source of qualitative data concerning our participants' professional experiences as well as a name generator for their support networks. Some questions were drawn from prior studies of support networks from an "exchange" perspective,[56] adapted for workplace contexts. Others used a more "role" perspective not centered on specific exchanges, such as asking about mentor figures. We used a semi-structured approach in the interviews, meaning interviewers would follow a protocol but would change question order, ask additional follow up questions, or change the language of questions in real time to allow for a natural conversation pattern and to attend to the overall goals of the interview.[57] The full interview protocol is available in Appendix A.

[55] Melinda McCormick, Ramon Barthelemy, and Charles Henderson, "Women's Persistence into Graduate Astronomy Programs: The Roles of Support, Interest, and Capital," *Journal of Women and Minorities in Science and Engineering* 20, no. 4 (2014), https://doi.org/10.1615/JWomenMinorScienEng.2014009829.

## Sociograms

During the interviews, we asked participants to add people to sticky notes on a Google Jamboard as they came up in conversation. Unfortunately, at the time of writing, Google Jamboard has been discontinued, though we were able to archive all sociograms as PDFs. A blank Jamboard (sociogram template) is shown in Figure 2. This template is based on one described by Antonucci[58] and our approach to using it is based on guidance from Hollstein and collaborators.[59] Typically, the interviewer would explain the intended use of the Jamboard at the beginning of the interview, and some participants would add names to the Jamboard as they spoke without prompting from the interviewer, while some would do so after prompting. Some also placed the sticky notes on the concentric circles of the Jamboard, while some left them to the side. Either way, at the end of the interview, the interviewer would ask the participant to arrange the sticky notes on the concentric circles according to closeness, which participants were sometimes asked to define in their own words. This typically meant that participants would place people they were closest to (often partners, family, and close friends and colleagues) near the star at the center that represented the participant. Their definitions of closeness were varied and interesting in their own right because they reveal which qualities delineate the "inner circle" from the rest of their network. A forthcoming study from this project looks specifically at this feature of our methods and how participants defined closeness, and preliminary results suggest that trust, relaxed boundaries, reliance, and support are especially important markers of close relationships.

---

[56] Mart G. M. van der Poel, "Delineating Personal Support Networks," *Social Networks* 15, no. 1 (March 1, 1993): 49–70, https://doi.org/10.1016/0378-8733(93)90021-C.
[57] "Qualitative Inquiry and Research Design | SAGE Publications Inc," accessed May 13, 2024, https://us.sagepub.com/en-us/nam/qualitative-inquiry-and-research-design/book246896.
[58] Toni C. Antonucci, "Measuring Social Support Networks: Hierarchical Mapping Technique," *Generations: Journal of the American Society on Aging* 10, no. 4 (1986): 10–12.
[59] Betina Hollstein, Tom Töpfer, and Jürgen Pfeffer, "Collecting Egocentric Network Data with Visual Tools: A Comparative Study," *Network Science* 8, no. 2 (June 2020): 223–50, https://doi.org/10.1017/nws.2020.4.

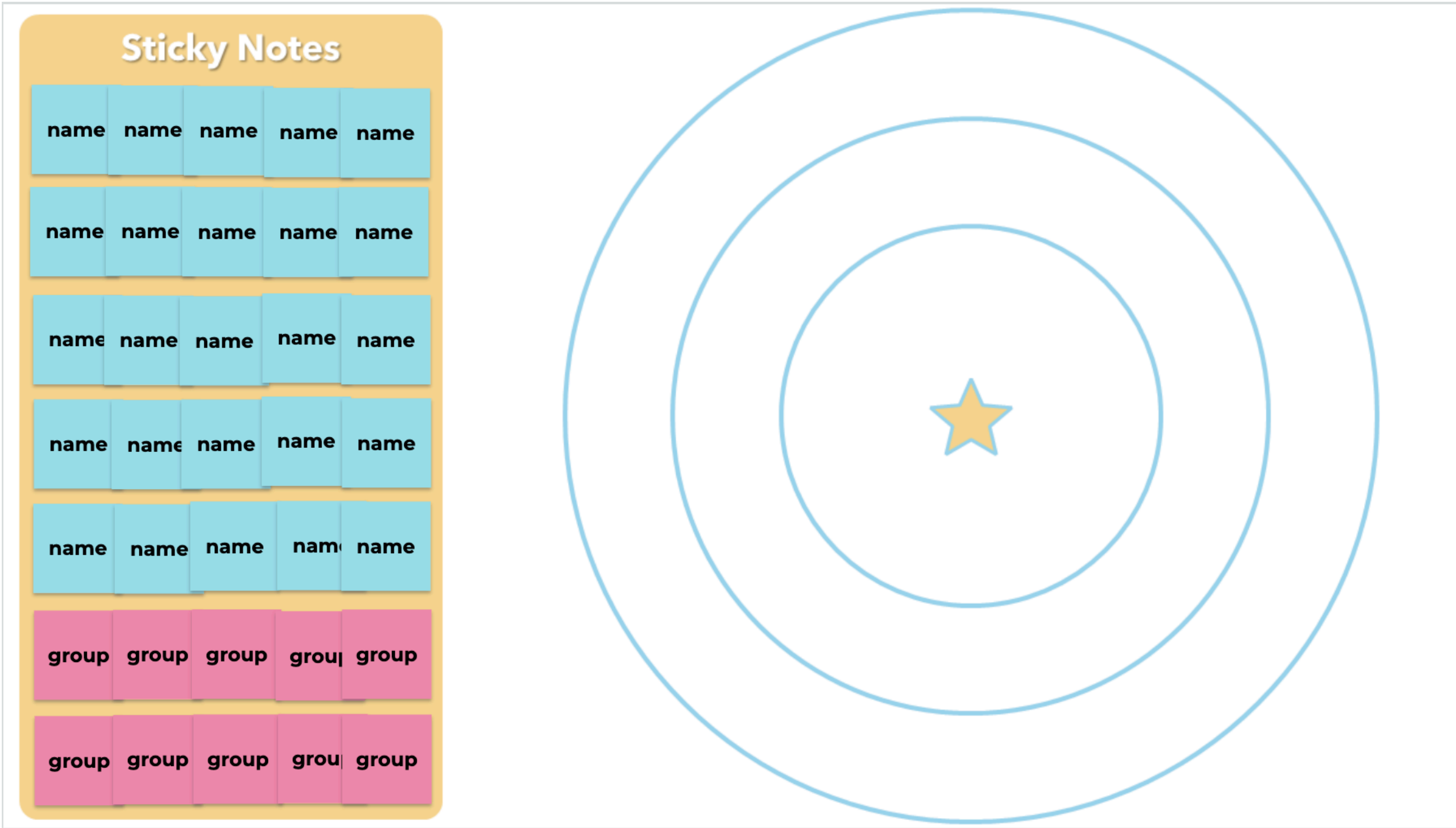

Figure 2: Sociogram template as Google Jamboard.

The interviewer would also ask participants to indicate which of their alters knew each other by drawing lines (using a built-in drawing tool) connecting them or closed loops around them indicating groups of people who knew each other.[60] Again, participants were left to determine what constituted "knowing" someone to themselves. With the alters placed on the Jamboard and connections among alters indicated, the Jamboards serve as sociograms (visual representations of social networks). Our interviews typically concluded with the interviewer asking for demographics like race, nationality, gender, and sexuality about each alter. Sometimes, usually due to time constraints, this question was left out, or our participants answered it in varying ways such that we do not have complete demographic information for many of the alters. A fictionalized example of a completed sociogram is shown in Figure 3.

---

[60] Bernie Hogan, Juan Antonio Carrasco, and Barry Wellman, "Visualizing Personal Networks: Working with Participant-Aided Sociograms," *Field Methods* 19, no. 2 (May 1, 2007): 116–44, https://doi.org/10.1177/1525822X06298589.

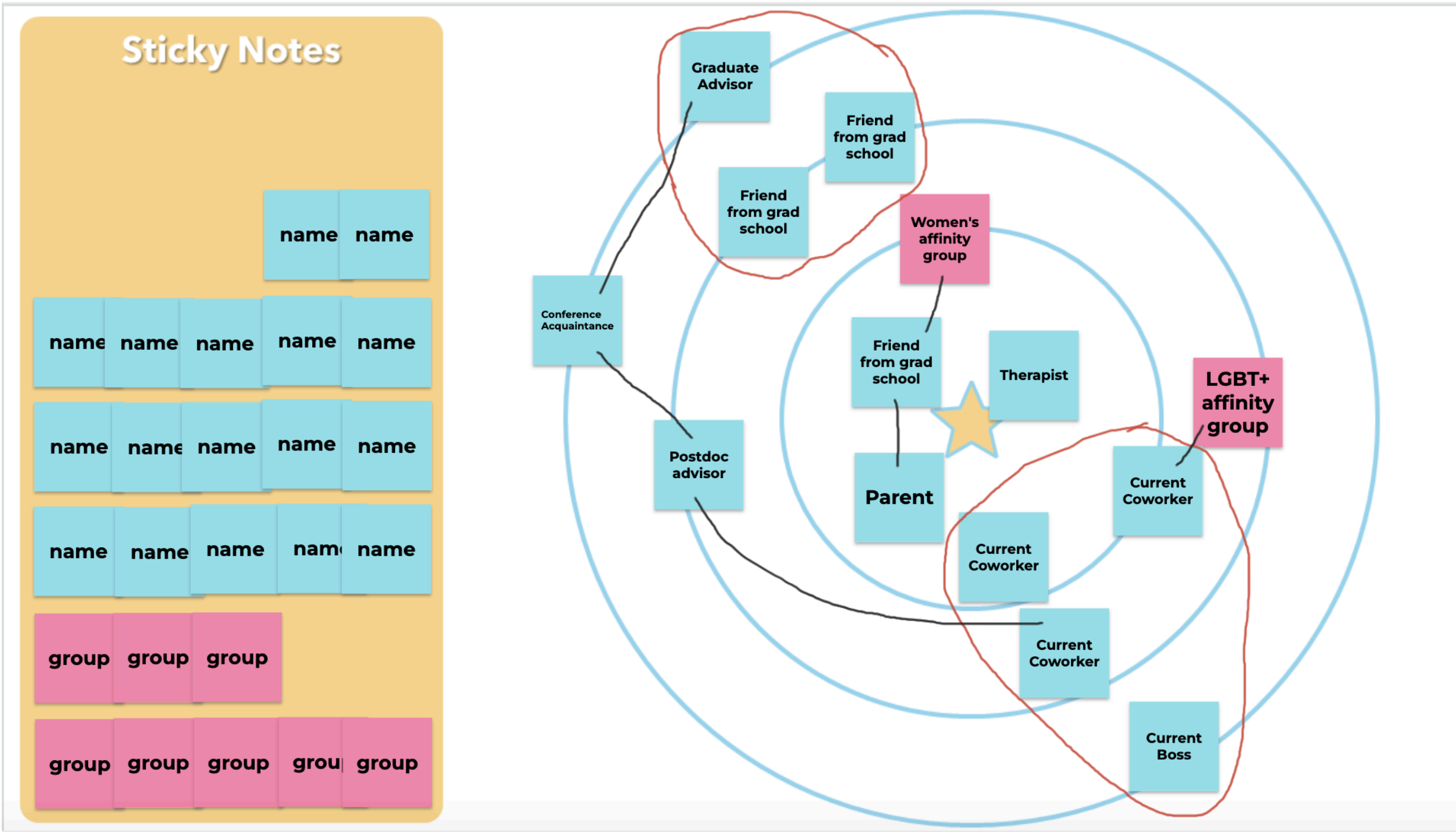

Figure 3: Fictionalized example of a sociogram showing alter connections.

## B. Thematic Coding

We used an emergent thematic coding scheme based on constructivist grounded theory,[61] meaning that we identified and marked common themes in the transcripts during initial reads, comprising the initial codebook. We further refined the codebook over the course of subsequent reads and discussions, keeping in mind the theoretical framework of professional support we were primarily investigating.  The coding team (Gutzwa, Swirtz, Manni, and Traxler) coded a subset of interviews multiple times to establish interrater reliability and continued to discuss and refine the codebook as they went. They used MAXQDA, which is a software program designed for working with qualitative data.

As part of the qualitative transcript coding, the team included two first level code categories related to network data in the interviews: Connection Nodes and Nodal Support. As participants mentioned other people or groups relevant to their professional journeys, they would be coded as a Connection Node with a corresponding subcode and sub-subcode indicating one or more affiliations and/or relationships to the participant ("Collegial & Supervisorial Codes" > "coworkers/peers," e.g.). When relevant, these mentions of alters would also be coded with types of support they are associated with or a gap in support, which did not have an associated type but which often had a type indicated by other Nodal Support codes. In this way, the qualitative coding process, which has and will continue to serve other studies related to this project, also serves as the starting point for the network data coding. All code categories and sub-categories, with network data codes emphasized, are listed in Table 1.

[61] Kathy Charmaz, "Constructivist Grounded Theory," *The Journal of Positive Psychology* 12, no. 3 (May 4, 2017): 299–300, https://doi.org/10.1080/17439760.2016.1262612.

Table 1: Codes, subcodes, and sub-subcodes used for coding interviews. Codes leading to network data are bolded.

| **1st level** | **2nd level** | **3rd level** |
|---|---|---|
| **Connection Nodes** | **Individual** | **Partners**<br>**Friends**<br>**Family**<br>**Other** |
| | **Group** | **Subfield**<br>**Group/listserv**<br>**Conferences**<br>**Affinity group** |
| | **Institutional** | **Undergraduate program**<br>**Graduate program**<br>**Postdoc**<br>**Past job**<br>**Current job**<br>**Future job** |
| | **Collegial & Supervisorial** | **Bosses/supervisors**<br>**Mentors (professional)**<br>**Mentees/direct reports**<br>**Coworkers/peers** |
| **Nodal Support** | **Identity-based support**<br>**Physical closeness**<br>**Networking support**<br>**Community building support**<br>**Career advice support**<br>**Emotional support**<br>**Material support**<br>**Instrumental support**<br>**DEI initiatives/policies**<br>**Gaps in support** | |
| Career Movement | Job Movement | Searching for jobs<br>Being offered jobs<br>Pulling into job<br>Pushing out of job<br>Pros of job<br>Cons of job |
| | Sector Movement | Pulling into sector<br>Pushing out of sector<br>Pros of sector |
| | | Cons of sector |
| Individual Values | Sector Comparison | Academia<br>Industry<br>Government |
| | Identity-based Values | Self-concept<br>Professional/academic identities<br>Social Identities |
| | (Inter)personal Values | Family needs<br>Personal goals/values<br>Location-based values |
| | Sector Values | Positive sector culture<br>General sector culture<br>Negative sector culture |
| | Career Values | Thoughts on changing/staying in careers<br>Would (not) recommend career path to others |
| | Functional Values | Job responsibilities<br>Being professor/teacher<br>Being physicist/researcher<br>Being mentor/supervisor |
| Secondary Analysis | 11 codes, including: "COVID," "Good quotes," "outness," and "explicit discrimination" | |

## Network Data Coding

After the initial round of coding, which included all of the code categories in Table 1, our team developed a secondary coding scheme for network data. The goal of this second coding round was to generate quantitative network data from the qualitative network data that was coded for in the interview transcript and to associate network data with named alters. We were interested in keeping track of each time an alter was mentioned for two reasons: (1) to have some quantifiable measure of importance or prominence and (2) to keep track of the qualitative content of each mention of an alter (as in, to know what type of support that specific mention was associated with, or which professional relationship was salient in that mention). This round of coding took place in a spreadsheet (the "Nodes" sheet) in which each mention of an alter was added as a new line with checkboxes indicating all relevant relational (Connection) and Nodal Support codes and the transcript line number. We added an additional checkbox called "Dissonance," which we marked when the mention included some indication of interpersonal tension, friction, or a feeling of "I don't like them," since we found that such feelings were important but not always captured by our Nodal Support codes.[62] We also specified types of gaps in support since those were not explicitly marked in the first round of coding.

The two coders for the network data (Hatcher and Donis) coded some interviews together to establish reliability, but generally engaged in little interpretation of the data since the network data were already coded in the first round—they were primarily associating those codes with specific alters. Still, there were cases where the combination of interview transcript, audio recording, and screen recording of the Jamboard left the identity of an alter being discussed or which codes belonged to which alter ambiguous, and in those cases the coders met to discuss and assign alters to codes. We discuss such cases in more detail in Section IV. C.

The secondary round of coding also included recording connections between alters in a spreadsheet (the "Connections" sheet). This took form as a two column list of names in which each pair of side-by-side names was connected—an edge list. These connections came from the sociogram itself, since participants were asked to indicate which of their alters knew each other. When multiple alters were defined as a group (usually by a closed loop drawn around them), we recorded connections between each group member. Each ego was also included in this spreadsheet as being connected to every alter in their network. All of this was done to prepare our data for input into SNA programming environments that typically take two-column lists of names as connections data inputs.

After this secondary round of coding was complete, we further manipulated the Nodes sheet to make it more easy to import into SNA software, which typically require lists of nodes with all attributes in additional columns. In our case, this meant making a copy of the Nodes sheet and consolidating all mentions of each alter (rows in the spreadsheet) down to one line which contained all of the codes associated with that alter. For example, if Alter A was coded as a Boss/Supervisor associated with a gap in material support in one instance and then as a Mentor associated with career advice support in another instance, those codes (along with any other codes that alter may be associated with) would all be recorded on one line so that Alter A would show up in that ego's network as a Boss/Supervisor *and* Mentor who was associated with both career advice support *and* a gap in material support.

This consolidation procedure is one of the clearest examples of the way that our SNA approach reduces complexity in our dataset in order to conduct meaningful quantitative analysis. We lose information about the way that the alters' associated codes appear in different contexts in the interviews, so our networks are ignorant of whether the various professional relationships or types of support our alters are associated with became associated with them all at once, in different conversational settings, in different temporal settings, or something else. We see this as a necessary simplification to make, since we want to examine alters holistically (across time, context, and ego affect), and since our interview and coding processes did not attempt to account for an exact and linear professional timeline for our participants (this was an intentional decision partially based on theories about differences in the way queer people experience time, especially around normative life progressions).[63] We also retained the original spreadsheet which includes line numbers for each separate mention, so the more detailed information can be used in future analyses.

[62] Peter V. Marsden, "Network Data and Measurement," *Annual Review of Sociology* 16 (1990): 435–63.

## Additional Quantitative Parameters

We created a parameter called "Mentions" to track the number of times an alter was mentioned. We distinguished between mentions of alters that were prompted by the interviewer versus those that happened as the result of a less direct question. For example, a mention of a PhD advisor that followed an interviewer statement like, "Tell me more about your PhD advisor" would be marked as a prompted mention, whereas one that followed a question like, "who did you go to for help in your PhD program?" would not. Ultimately, we decided to count prompted mentions as one half of an unprompted mention when calculating total Mentions for each alter. This was based on the notion that an unprompted mention represented slightly more prominence, importance, or impact, but that a prompted mention still represents some, even if it may be filtered through the researcher's interview process. Prompted mentions of alters are an example of how researchers (interviewers) can shape our data, since the interviewer makes a decision about which alters to ask further questions about, but we see it as part of a tradeoff in using semi-structured interview protocols (the benefits being the natural and flexible mode of conversation). We do not think the decision to reduce prompted mentions to one half of an unprompted mention is the only reasonable approach to measure mentions, especially since it flattens their varied nature, but we see this also as part of a tradeoff in the analytical power of having a single parameter to use in analysis. We also retain the line-by-line coding of mentions (and whether they were prompted or not), so the data is not lost, just simplified in the parameter.

Additionally, we added a parameter called "Proximity" that quantified the placement of an alter on the sociogram in relation to the ego (represented by the star in the middle). We used the concentric circles on the sociogram as reference points to assign integer values for proximity ranging from 1-8 based on the location of the center of each sticky note. A diagram showing this coding scheme is given in Figure 4. Occasionally, study participants left alters in the sticky note field rather than placing them on the diagram. This could have been due to interview time constraints, fatigue involved in constructing large networks, or uncertainty about how to assign position to those alters—we are not sure. In those cases, we left "Proximity" for those nodes blank, which became "N/A" in most coding environments. We decided to call this parameter "Proximity" instead of "Closeness" because, even though our participants were asked to arrange their alters according to closeness, closeness has a reserved meaning as a measure of centrality in SNA contexts and our participants had varying definitions of what it meant to them.

[63] Elizabeth Freeman, *Time Binds: Queer Temporalities, Queer Histories* (Duke University Press, 2010), https://doi.org/10.1215/9780822393184.

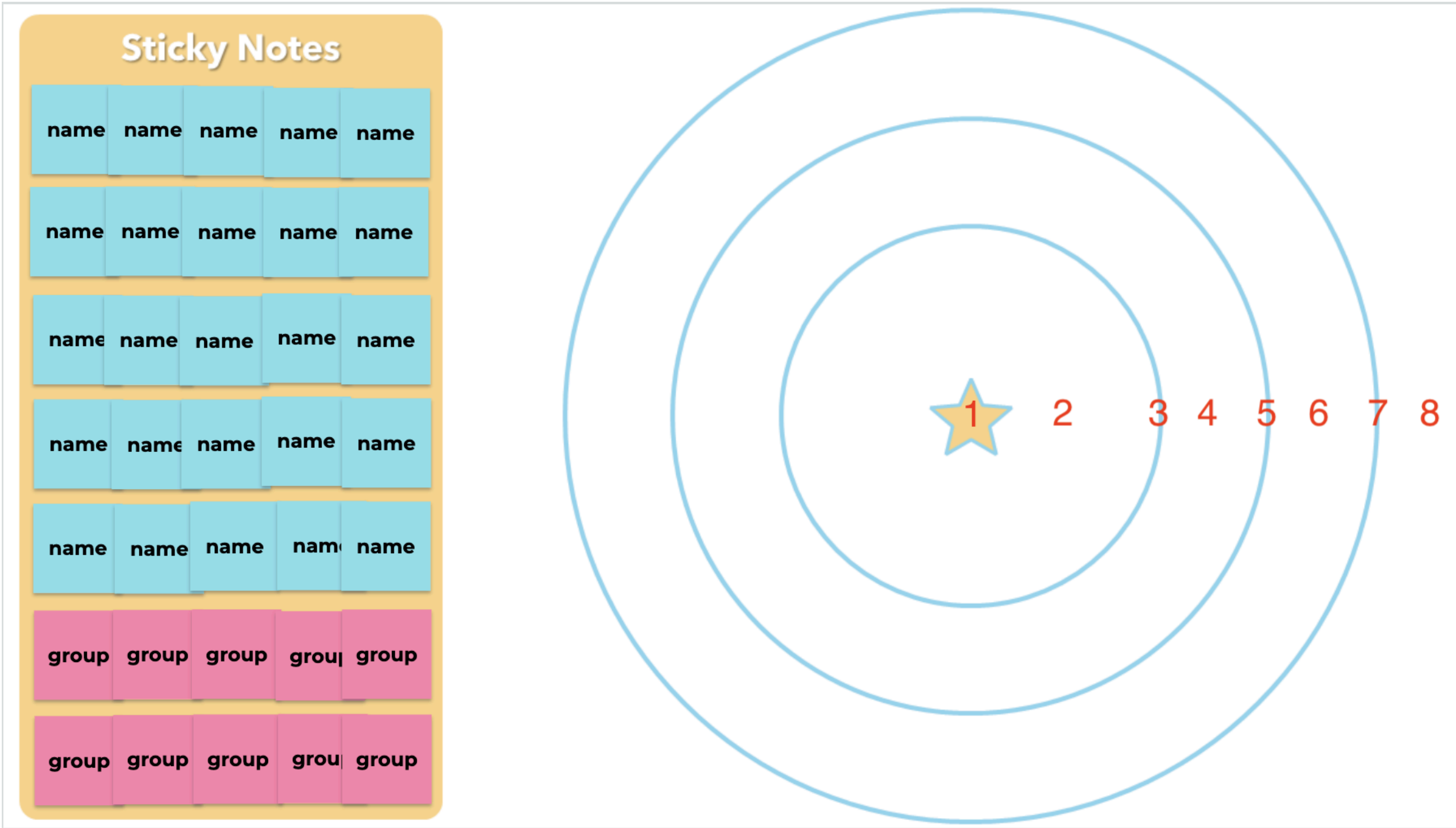

Figure 4: Sociogram template with numbers indicating Proximity values assigned to sticky notes placed at those radial positions.

## C. Computational Analysis

In this section, we discuss and explain the computational analysis included in our methods, all of which we coded in R.[64] A simplified and less project-specific version of our code, along with anonymized sample data, is available in the Supplemental Material.

### Data Preparation

The consolidation discussed in the preceding section could be seen as front-end data preparation. This was done by hand partly because of the user-friendly nature of spreadsheets for data manipulation and partly so that any errors or notes related to anomalous events in the spreadsheet could be addressed. The back-end data preparation came after the data were downloaded as .csv files and uploaded in the programming environment, which in our case was Rstudio.[65] Those data inputs are as follows:

1. ***Connections:*** 3 column matrix with interview number (also functioning as ego identifier), Name 1, and Name 2 as columns. Name 1 and Name 2 are two nodes to be connected in a network, including alter-alter and alter-ego ties. This file contains all connections data for all interviews.

[64] R Core Team, *R: A Language and Environment for Statistical Computing* (Vienna, Austria: R Foundation for Statistical Computing, 2020), https://www.R-project.org/.
[65] R Studio Team, "RStudio: Integrated Development for R" (Boston, MA: PBC, 2020), http://www.rstudio.com/.

2. ***Attributes:*** contains Name, associated interview number, and all attributes (Connection Nodes codes, Nodal Support codes [with separate columns for each code in these categories], Proximity, Mentions, as well as personal demographics for some) as separate columns. Rows are individual alters.
3. ***Ego Attributes:*** contains Name, associated interview number, and all attributes (professional and personal demographics including employer, employment sector, gender, and race) as separate columns. Rows are study participants.

## Metrics to Calculate

We were interested in the metrics that revealed characteristics of our study population's networks, especially in the way these values were distributed across our study population. Visualizing these distributions is both informative and exploratory because it immediately gives an idea of what a measurement of a given characteristic looks like for the whole population as well as potentially giving direction for further inquiry, especially if we find non-normal or otherwise unexpected distributions. A table of the metrics we calculated and their explanations can be found in Table 2.

Table 2: Characteristics and metrics describing whole networks and their associated definitions.

| **Network characteristic or metric** | **Description** |
|---|---|
| Total Nodes | Number of alters or groups |
| Total Groups | Number of groups |
| Percent Groups | Percent of Total Nodes which are Groups (Total Nodes divided by Total Groups) |
| Total Dissonance | Number of nodes carrying code for "Dissonance" |
| Percent Dissonance | Percent of Total Nodes carrying code for Dissonance (Total Nodes divided by Total Dissonance) |
| Average Proximity | Mean of proximity measures |
| Average Mentions | Mean of mentions |
| Total Support | Number of instances of support (each node could have up to 9 associated types of support, see Tab. 1) |
| Total Gaps in Support | Number of instances of gaps in support (each node could have up to 9 associated types of gaps in |
| | support) |
| Support to Nodes Ratio | Ratio of Total Support to Total Nodes |
| Gaps in Support to Nodes Ratio | Ratio of Total Gaps in Support to Total Nodes |
| Density | Global density, or total number of connections divided by total number of possible connections (igraph's edge_density function) |
| Diameter | Largest separation between two nodes in the network as measured by number of linking edges (igraph's diameter function) |
| Components | Number of isolated nodes or groups of nodes (igraph's count_components function) |
| Largest Clique | Number of nodes in largest clique (fully-connected component, would be Total Nodes if Components = 1, found with igraph's clique_num function) |
| Cliques | Number of distinct cliques (fully-connected components) of three or more nodes (from igraph's max_cliques function; setting min = 3) |
| Transitivity | Global transitivity (sometimes called clustering coefficient), or the probability that adjacent nodes are connected (igraph's transitivity function) |

The metrics and characteristics in Table 2 do not include the ego as part of the network. Depending on research goals, it may or may not be appropriate to exclude ego from their network for the calculation of various metrics. Since the ego is necessarily connected to all other nodes in their network, they have at best a predictable and at worst a confounding effect on network metrics. McCarty and Wutich note that excluding the ego may be especially important for understanding network mechanics that do not involve them (gossip, for example), and that when included, the ego will typically dominate the network, so one would be studying the effect of the ego on the network rather than the other way around.[66] Because we are more interested in the effects of the network on the ego, and because of the potential to confound results (Diameter, for instance, is always 2 with ego included), we choose to remove ego from the network for these calculations. After calculating all metrics and characteristics in Table 2 for each network, we used histograms to visualize distributions and look for areas deserving further inquiry.

[66] Christopher McCarty and Amber Wutich, "Conceptual and Empirical Arguments for Including or Excluding Ego from Structural Analyses of Personal Networks," *Connections* 26, no. 2 (2005): 82–88.

### Categorization

In addition to metrics and characteristics that describe the entire networks, we were also interested in the composition of our participants' networks based on a few categorical identities of alters we could construct from our codes. One way to examine the composition of participants' networks is through the earliest career stage in which they discuss their alters. We created a new categorical alter attribute called “Stage” which indicates the earliest career stage with which that alter is associated. In this way, “Stage” represents a sort of “age” of a connection on an academic/professional timeline, but not necessarily the overall nature of the connection. The Institutional Connection Node category is most useful for this categorization, since the code types follow a generally linear (and widely adhered to) progression through a PhD program and towards a career (Undergraduate Program, Graduate Program, Postdoc, Past Job, Current Job, and Future Job). If an alter is coded as being associated with ego's Undergraduate Program, they would be listed as “undergrad” in the “Stage” attribute, regardless of whether they were also coded for Graduate Program or any other connection code. Other codes follow suit, except for Past, Current, and Future Job. These were all coded as “job,” since the amount of time spent in any of these positions would be more variable than the academic positions, and since they would all include the professional career of the ego. If alters did not have codes for the Institutional Connections category (e.g., family members), their “Stage” was categorized as “none.”

We also were interested in examining network composition based on a general relationship type. We wanted to do this because the 18 Connection Node code types were too compartmentalized and overlapped too much for intuitive comparison. Instead, we added the attribute “Relationship” for each alter and assigned categories according to groupings of the 18 code types and based on common relationships. For example, the relationship category “Groups” would be assigned to any alter coded as Group/listserv and/or Affinity Group and nothing else. We ended up with 8 Relationship categories: Peer, Mentor, Groups, Friend (professional), Friend, Family/partners, Mentees, and Boss. The “Relationship” category, unlike “Stage,” does try to represent an overall nature of a connection.

## D. Network Visualization

Visualization is one of the primary tools for analyzing networks. Different layout algorithms make certain features especially visible, and color, shape, and other visual cues can encode information. We used a combination of arrangement and encoding to highlight various features of our networks. Visualization techniques represent the bulk of the code we have made available as Supplemental Material.

### Standard Visualization

We use this term to refer to direct visualizations of our networks that include all nodes and connections without additional rearrangement or consolidation. We used three main R packages to produce standard visualizations: igraph,[67] tidygraph,[68] and egor.[69] The first two are widely used for general network analysis while the third is for egocentric analysis and visualization specifically.

Using igraph and tidygraph, we generated standard visualizations that encoded information about Proximity and Mentions using node color and size, respectively. We were also able to encode information about support (or gaps therein) in the connections between alters, with color representing the support or gaps in support (green and red, respectively, or black for none) and edge thickness representing number of different support types. An example is shown in Figure 5. We represented the networks this way with the ego both included and not included, though in the ego-less representations, support is also not represented.

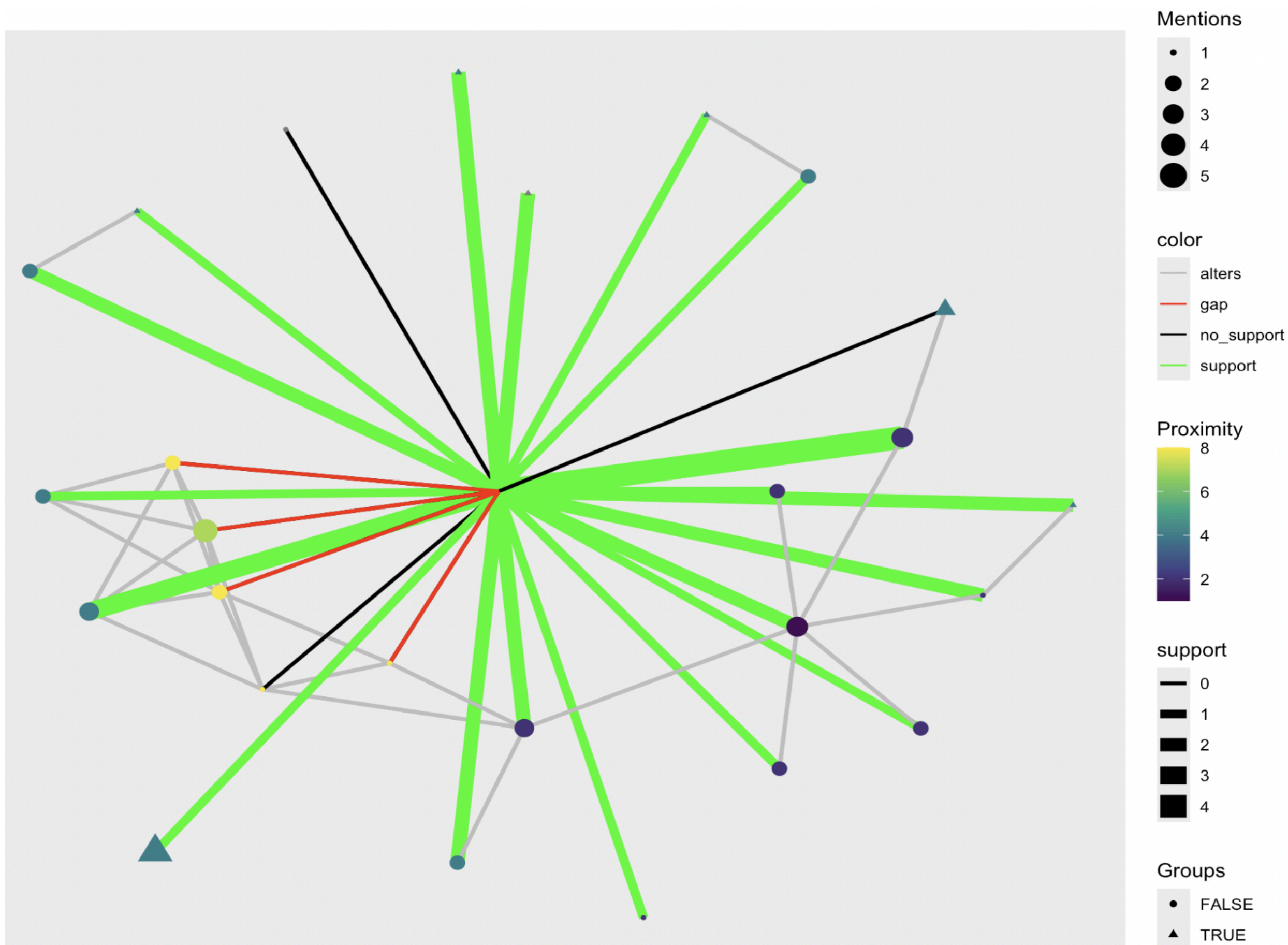

Figure 5: Standard visualization made using tidygraph and igraph encoding Proximity and Mentions as node color and size, respectively, presence of support or gaps in support as edge color, and number of types of support as edge thickness.

[67] Gabor Csardi and Tamas Nepusz, "The Igraph Software Package for Complex Network Research," *InterJournal* Complex Systems (2006): 1695.
[68] Thomas Lin Pedersen, *Tidygraph: A Tidy API for Graph Manipulation*, 2024, https://tidygraph.data-imaginist.com.
[69] Till Krenz et al., "Egor: Import and Analyse Ego-Centered Network Data," July 2, 2018, https://doi.org/10.32614/CRAN.package.egor.

We used egor to represent the networks in the standard way also, though we chose to do so in a way that does not include additional information (like Mentions and Proximity of nodes) partly because we already created such representations using igraph and tidygraph and partly because egor's default visualization scheme lent itself well to comparison by creating more uniform arrangements. An example of a few sociograms is shown in Figure 6 to highlight their comparability.

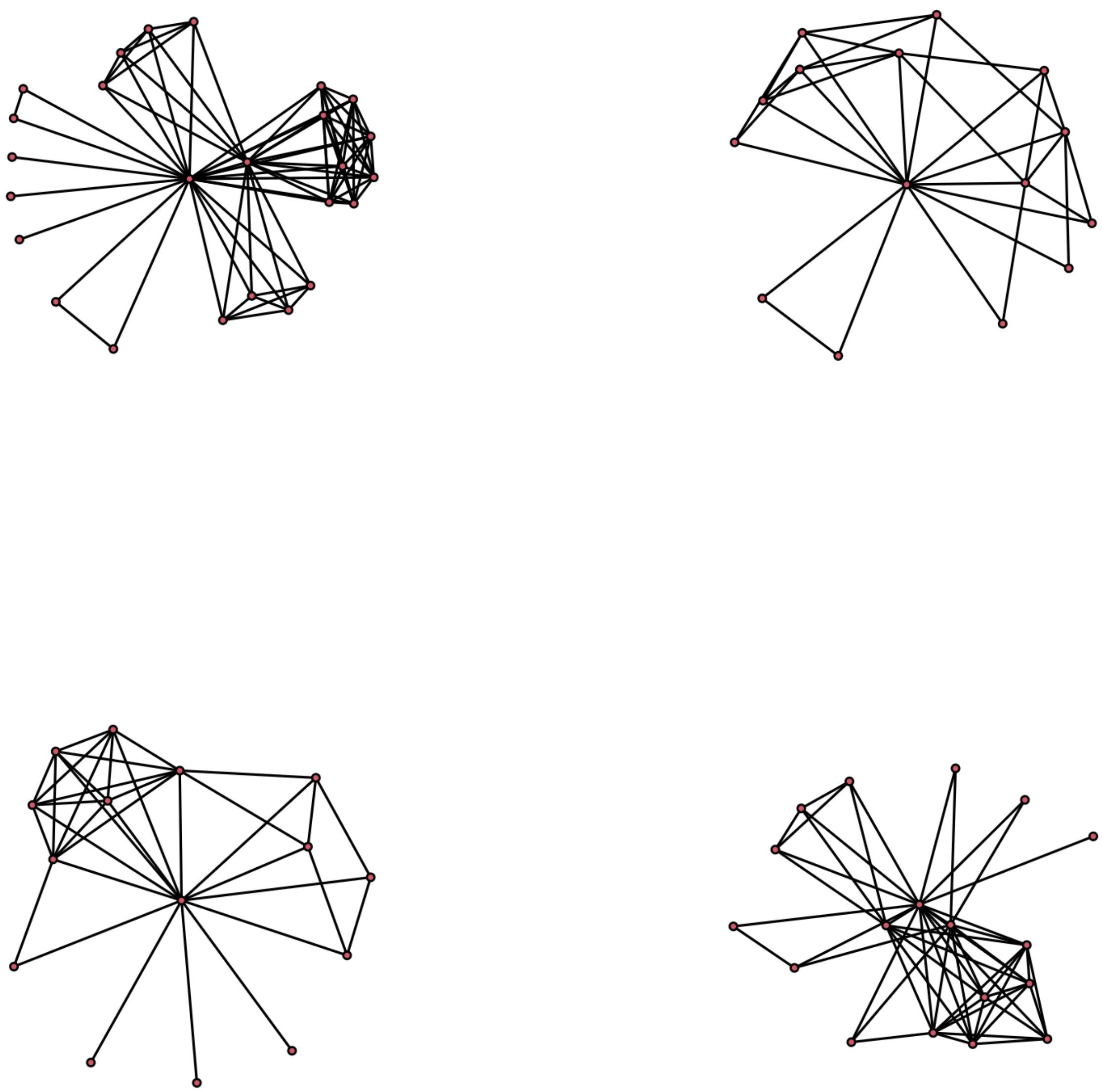

Figure 6: Standard network visualizations made with egor, highlighting the uniformity of the arrangement algorithm.

## Simplified Visualization

In addition to examining individual networks, we were interested in comparing across career sectors or other dimensions of personal identity. To do this visually, it was helpful to simplify our networks to a similar basic and spatially simple structure, since comparing 100 networks by eye would be too challenging or time consuming. One method was to collapse similar alters into one node and let variable size and color describe these groups of nodes. This was where our categorizations discussed above (Stage, Relationship, etc.) could be especially useful. Figure 7 shows an example for the Stage parameter, where size shows number of nodes, color shows connectivity within the group, and connecting edge thickness represents connectivity outside/with another group. The egor function "clustered_graphs"[70] performs this consolidation and plots the results. See the cited paper or other tutorials and guides for using egor[71] for a more thorough explanation of how to use this function (and others), or the Supplemental Material for an example of how we used it. This is an example of simplification through data reduction: some information, specifically the actual connections among alters, is lost by visualizing the networks this way. We think this is a reasonable tradeoff to have an easier basis of comparison, especially considering the information lost is accounted for in other parts of this analysis.

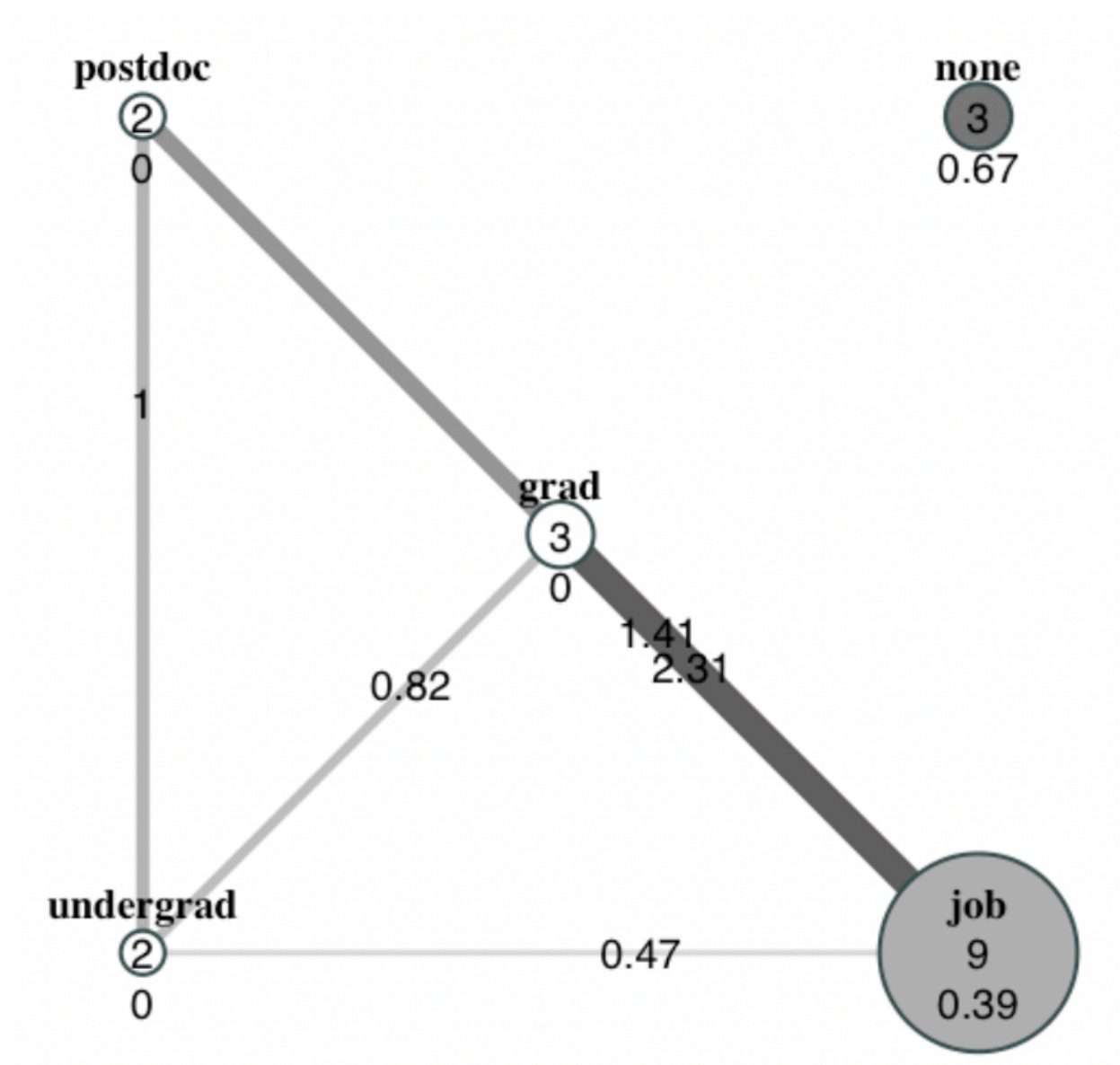

Figure 7: Simplified clustered graph visualization with "stage" as clustering parameter. Node size indicates number of nodes having that classification, node darkness indicates connectivity within that classification, edge thickness and darkness indicate connectivity between two classes.

We also compared networks by assigning spatial regions to categorical attributes without combining nodes. Though this mode of visualization included all nodes and connections, we consider it "simplified" because it forces a certain uniform spatial arrangement on all networks for the sake of comparison. We used a plot field of concentric circles where different regions of the field are associated with different categories. For example, to arrange alters based on Relationship, the field was divided into 8 equal-sized sections with alters placed appropriately. In this case, connections were still shown, but we also represented numerical node attributes (like Mentions) by node size and Proximity by closeness to the center of the diagram (much like our participants' original sociograms). The egor function "plot_egor" (with "type" set to "egogram" as described in Krenz's tutorial[72] and in the Supplemental Material) resulted in a diagram as shown in Figure 8.

[70] Ulrik Brandes et al., "Visual Statistics for Collections of Clustered Graphs," in *2008 IEEE Pacific Visualization Symposium*, 2008, 47–54, https://doi.org/10.1109/PACIFICVIS.2008.4475458.
[71] Till Krenz, "Using Egor to Analyse Ego-Centered Network Data," February 1, 2024, https://cran.r-project.org/web/packages/egor/vignettes/using_egor.html.

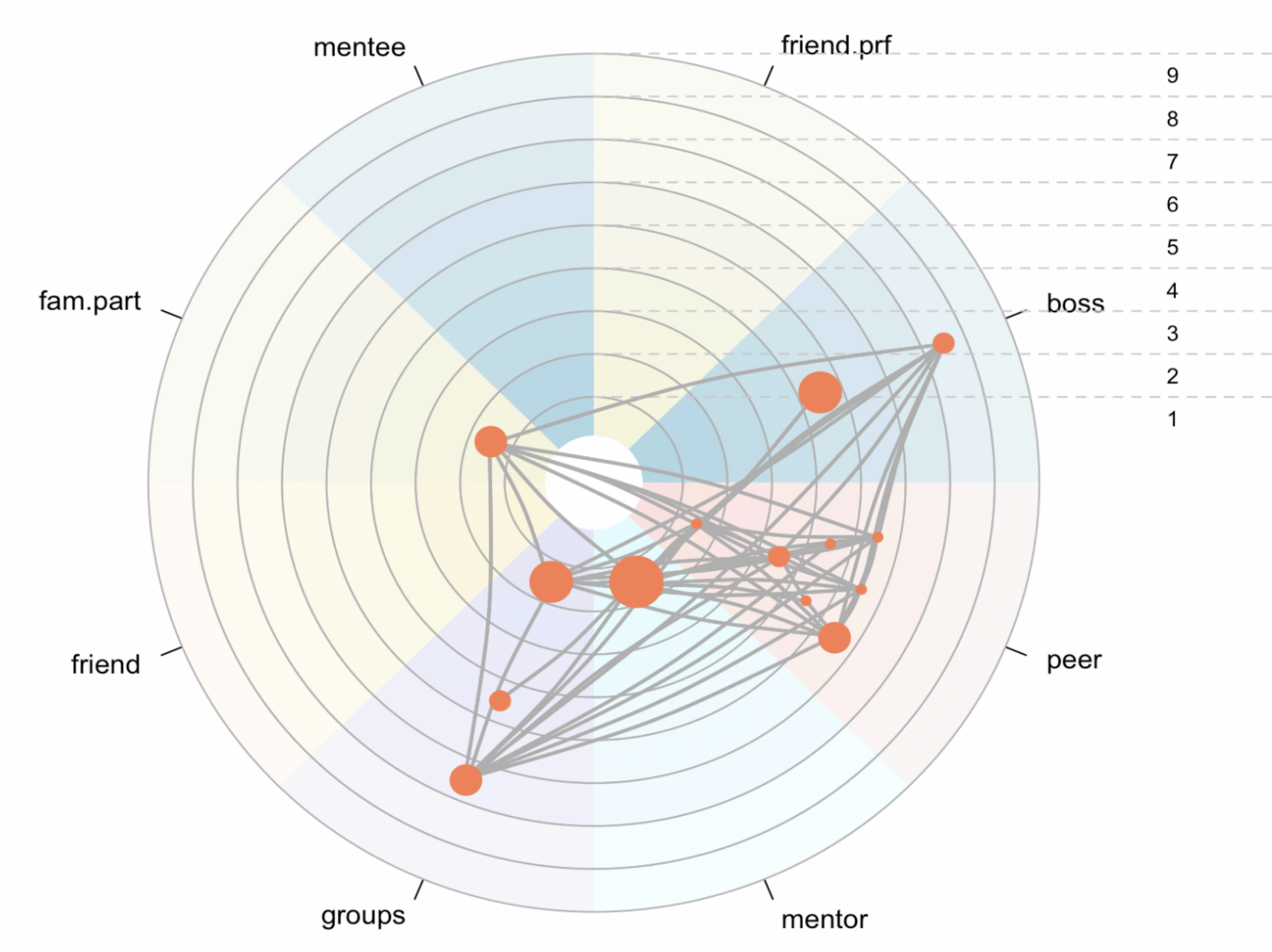

Figure 8: Egogram with alters grouped according by Relationship, sized by Mentions, and organized radially by Proximity. Note: this function requires the variable assigned to radial position to be numeric, so Proximity measures of N/A were changed to 9 when creating these diagrams.

[72] Krenz.

# IV. Discussion

We have presented a somewhat novel set of methods for quantitative egocentric network analysis based on qualitative data. We have considered critical methodologies and chosen SNA modalities accordingly with the hope that our study can be as attentive as possible to the nuance and complexity of our participants' experiences as marginalized physicists. In this section, we discuss the novelty, limitations, and potential applications of the methods. We conclude with a discussion of future work and some closing remarks.

## A. Novelty

In our review of SNA literature, we have not found a study that performs egocentric network analysis the way we have in this project. Many aspects, like name generator structure, in-interview sociogram construction, and the focus on support as a basis of connection are not new to SNA,[73] but the way we have combined these components is new. Similarly, the egor library is not our creation, but it has not been used in PER as far as we are aware.

Some aspects of our methods are completely new, and it is worth emphasizing these aspects so that those wishing to adopt our methods understand the areas that are not well explored. As far as we are aware, researchers have not quantified and coded "Mentions" as we have in this study. By tracking each mention of an alter separately, not only do we have a more detailed record of the way they relate to and support the ego in different instances in the interview, but we also believe we have some estimate of the alter's prominence or consequence in the eyes of the interviewee. Our theory is that, in a long discussion of one's professional journey (like one of our interviews), a person that is brought up often in conversation likely has some significance in that journey, though the type of significance would depend on the context in which they are brought up. More research would be needed to understand exactly how the number of alter mentions relates to that alter's relationship to the ego, but we suggest that it represents something related to consequence or importance in the ego's network. At the very least, it serves as a way to measure how much an alter is discussed in the interview, which is an interesting metric on its own. Our use of a "Proximity" parameter to quantify placement of alters on the concentric circles of our sociogram template is also new, and, like "Mentions," seems to represent some type of centrality of alters (literally "closeness," but as defined by egos, not in the traditional network sense). We invite other researchers to use "Mentions" and "Proximity" as metrics in similarly-designed studies and report their experiences and results.

Our use of the egor package, specifically its simplified network visualizations, is new to the field of PER, as far as we are aware. The simplified network visualizations could be very useful for comparing large numbers of egocentric networks at a glance or perhaps for displaying artificial "average" networks for a certain class of egos (see Brandes 2008).[74] Since this package is tailored for use in egocentric analysis, we encourage other researchers designing egocentric studies to use its tools and report on their techniques.

[73] Antonucci, "Measuring Social Support Networks."
[74] Brandes et al., "Visual Statistics for Collections of Clustered Graphs."

## B. Limitations

This study was primarily limited by a tradeoff between faithful and complete representation of our qualitative data and the simplification of data needed for quantitative analysis. We believe it was worth simplifying and therefore reducing our data in order to perform this analysis, especially since we retained all qualitative data to support or refute quantitative conclusions and since the rest of our team is engaging in mostly qualitative analysis of our data. Still, we acknowledge that the present work will inherently involve data that are manipulated from their original form and therefore may contain some researcher bias.

We also note that some expected variation in interview and coding approaches among our team may have resulted in differences in the way that networks are described and constructed. For example, some interviewers wrote names of alters on the Jamboard for the participant, while some would ask the participant to write the names themselves. We have no reason to believe that this led to differences in the networks, but it is an instance in which researcher behavior may have impacted study results. This is to be expected with any qualitative (and some quantitative) data collection that involves a team of researchers.

In rare instances, researchers involved in the network coding were unable to determine which alters our participants were speaking about. We had video/audio recordings of the interviews, which often included live recording of the Jamboard where participants added alters, but occasionally there was not enough context for our coders to uniquely identify alters. In these cases, our coders met to discuss and tried to make the best choice with the information available. In the very rare cases where it was impossible to figure out who the interviewee was talking about, we did not code network data for those mentions.

Working with small and niche coding libraries, like egor, while very useful in specific applications, comes with drawbacks. We had some difficulty in modifying some of the display techniques, like adjusting colors and adding descriptive titles to figures. A combination of limited documentation and online discussion on some functions and maintenance of the libraries by small groups or individuals meant that troubleshooting was difficult and sometimes impossible. Overall, this amounted to minor and mostly cosmetic issues in the analysis, but it is worth noting for anyone planning to use highly specific coding tools.

Finally, this study was limited by features of SNA and egocentric studies in general, some of which we have already discussed. We asked our participants to provide information about alter-alter relationships and demographics, and they may not have always been correct in the information they provided. The same goes for information they provided about their own relationships, though in that case, they were the best possible source of information. Using a combination of interview and sociogram construction as our name generator meant that interview time constraints and parameters of the Jamboard space could have impacted network construction, especially for our Mentions and Proximity metrics. In general, when choosing to collect qualitative data and network data in an interview setting, we trade uniformity in the data collection process for richness and detail in the data, which we see as a worthwhile trade.

### C. Potential Applications

We see this set of methods as useful for anyone designing an SNA study that wants to address questions related to diversity, equity, or inclusion. It is not the only appropriate way to attend to such questions, but we designed it to be maximally attentive to the complexities of personal experiences with marginalization. In PER, this might mean questions related to belonging, identity, support, or success among diverse groups of physicists or physics students.

Specifically, this set of methods could be useful in studying the effects of DEI initiatives and programs which have been the subject of much scrutiny and debate lately. A comparison of the networks of physicists (or other professionals) that have or have not engaged regularly with such initiatives and programs would likely yield interesting results. This set of methods could also be useful in studying other professional programs and initiatives, especially if they are aimed at interpersonal connection. Analysis of personal professional support networks would be a great way to study the effects of a variety of policies, initiatives, interventions, and programs directed at professionals.

Our methods also provide useful examples for anyone designing MMSNA or qualitative SNA studies, as well as egocentric studies. As mentioned throughout this paper, we encourage researchers in PER to adopt these approaches to SNA, especially if they are interested in the effects of networks on individuals or on mechanisms in networks (beyond just descriptions of structure).

### D. Future Work

Future work in this project will involve the publication of a paper describing the results of this analysis, mostly focusing on comparisons of network characteristics across career sectors. Other work stemming from this analysis will follow, but the exact topics are not yet known and will likely come through exploratory analysis of our dataset.

Other goals of this project are the study of international participants, Participants of Color, disabled participants, and queer participants in our dataset. Members of our team are also investigating the themes and interconnected nature of different support types identified by participants, career values as described by participants, and qualitative comparisons of career sectors, among other topics.

## V. Conclusion

This project is based on 100 interviews with women and/or LGBTQ+ PhD-holding physicists about their career trajectories, professional journeys, and support networks. In this paper, we have presented a novel method for undertaking an egocentric network analysis of our dataset. We have noted how certain features of our methods, like the choice to use egocentric MMSNA, align with features of critical methodologies. We have also discussed the components of our methods which are completely new to the field of SNA, like the quantification of Mentions in interviews as a measure of consequentiality for alters.

Our methods could be adapted by anyone wishing to undertake a similar study, and we invite them to do so and share results of their work, and we encourage them to reach out to us if they would like to discuss these methods in more detail. Parts of these methods, especially in our data collection process and network data coding procedure, could prove useful to other researchers interested in egocentric SNA but not pursuing similar research. At the very least, our emphasis on critically-aligned techniques and our reporting on those techniques might be useful to anyone undertaking sociocultural studies in physics and wishing to attend to the complexities of identity, difference, and marginalization in their work.

The goal of this paper is to provide the PER community with a new example of how to approach SNA in a way that complements rather than strictly reduces qualitative data. We hope that other researchers will see the importance of grappling with questions of how their methods impact their results and their study participants, even if they are not undertaking SNA work. The ultimate goal of this work is to gain a better understanding of the way GSM physicists experience support so that we can advise institutions and professional societies on how to better support these physicists. Understanding of such a personal and complex phenomena as support requires careful planning and creativity in study design and thorough thoughtfulness in methods, which we have hopefully represented here.

# References

Antonucci, Toni C. "Measuring Social Support Networks: Hierarchical Mapping Technique." *Generations: Journal of the American Society on Aging* 10, no. 4 (1986): 10–12.

Atherton, Timothy J., Ramón S. Barthelemy, Wouter Deconinck, Michael L. Falk, Savannah Garmon, Elena Long, Monica Plisch, Elizabeth H. Simmons, and Kyle Reeves. "LGBT Climate in Physics: Building an Inclusive Community." College Park, MD: American Physical Society, 2016.

Barthelemy, Ramón S. "LGBT+ Physicists Qualitative Experiences of Exclusionary Behavior and Harassment." *European Journal of Physics* 41, no. 6 (October 2020): 065703. https://doi.org/10.1088/1361-6404/abb56a.

———. "Physics Education Research: A Research Subfield of Physics with Gender Parity." *Physical Review Special Topics - Physics Education Research* 11, no. 2 (2015). https://doi.org/10.1103/PhysRevSTPER.11.020107.

Barthelemy, Ramón S., Bryce E. Hughes, Madison Swirtz, Matthew Mikota, and Timothy J. Atherton. "Workplace Climate for LGBT+ Physicists: A View from Students and Professional Physicists." *Physical Review Physics Education Research* 18, no. 1 (June 13, 2022): 010147. https://doi.org/10.1103/PhysRevPhysEducRes.18.010147.

Barthelemy, Ramón S., Melinda McCormick, and Charles Henderson. "Gender Discrimination in Physics and Astronomy: Graduate Student Experiences of Sexism and Gender Microaggressions." *Physical Review Physics Education Research* 12, no. 2 (August 1, 2016): 020119. https://doi.org/10.1103/PhysRevPhysEducRes.12.020119.

Barthelemy, Ramón S., Melinda McCormick, and Charles R. Henderson. "Understanding Women's Gendered Experiences in Physics and Astronomy through Microaggressions," 35–38, 2014. https://www.per-central.org/items/detail.cfm?ID=13442.

Barthelemy, Ramón S., Melinda McCormick, Charles R. Henderson, and Alexis Knaub. "Educational Supports and Career Goals of Five Women in a Graduate Astronomy Program." *Physical Review Physics Education Research* 16, no. 1 (April 21, 2020): 010119. https://doi.org/10.1103/PhysRevPhysEducRes.16.010119.

Barthelemy, Ramón S., Madison Swirtz, Savannah Garmon, Elizabeth H. Simmons, Kyle Reeves, Michael L. Falk, Wouter Deconinck, Elena A. Long, and Timothy J. Atherton. "LGBT+ Physicists: Harassment, Persistence, and Uneven Support." *Physical Review Physics Education Research* 18, no. 1 (March 28, 2022): 010124. https://doi.org/10.1103/PhysRevPhysEducRes.18.010124.

Barthelemy, Ramón S., Adrienne L. Traxler, Jennifer Blue, and Madison G. Swirtz. "Research on Gender, Intersectionality, and LGBTQ+ Persons in Physics Education Research." In *The International Handbook of Physics Education Research: Special Topics*, edited by Mehmet Fatih Taşar and Paula R. L. Heron. AIP Publishing LLC, 2023. https://doi.org/10.1063/9780735425514.

Borgatti, Stephen P., Ajay Mehra, Daniel J. Brass, and Giuseppe Labianca. "Network Analysis in the Social Sciences." *Science* 323, no. 5916 (February 13, 2009): 892–95. https://doi.org/10.1126/science.1165821.

Brandes, Ulrik, Patrick Kenis, and Jörg Raab. "Explanation through Network Visualization." *Methodology* 2, no. 1 (January 2006): 16–23. https://doi.org/10.1027/1614-2241.2.1.16.

Brandes, Ulrik, Jurgen Lerner, Miranda J. Lubbers, Chris McCarty, and Jose Luis Molina. "Visual Statistics for Collections of Clustered Graphs." In *2008 IEEE Pacific Visualization Symposium*, 47–54, 2008. https://doi.org/10.1109/PACIFICVIS.2008.4475458.

Brewe, Eric. "The Roles of Engagement: Network Analysis in Physics Education Research." *Network Analysis*, n.d., 17.

Brewe, Eric, Laird Kramer, and Vashti Sawtelle. "Investigating Student Communities with Network Analysis of Interactions in a Physics Learning Center." *Physical Review Special Topics - Physics Education Research* 8, no. 1 (January 12, 2012): 010101. https://doi.org/10.1103/PhysRevSTPER.8.010101.

Bruun, Jesper. "Networks as Integrated in Research Methodologies in PER," 11–17, 2016. https://www.per-central.org/items/detail.cfm?ID=14295.

Carrington, Peter J, John Scott, and Stanley Wasserman. "Models and Methods in Social: Network Analysis," 2005, 345.

Celikates, Robin, and Jeffrey Flynn. "Critical Theory (Frankfurt School)." In *The Stanford Encyclopedia of Philosophy*, edited by Edward N. Zalta and Uri Nodelman, Winter 2023. Metaphysics Research Lab, Stanford University, 2023. https://plato.stanford.edu/archives/win2023/entries/critical-theory/.

Charmaz, Kathy. "Constructivist Grounded Theory." *The Journal of Positive Psychology* 12, no. 3 (May 4, 2017): 299–300. https://doi.org/10.1080/17439760.2016.1262612.

Cochran, Geraldine L., Theodore Hodapp, and Erika E. Alexander Brown. "Identifying Barriers to Ethnic/Racial Minority Students' Participation in Graduate Physics." In *2017 Physics Education Research Conference Proceedings*, 92–95. Cincinnati, OH: American Association of Physics Teachers, 2018. https://doi.org/10.1119/perc.2017.pr.018.

"Critical Race Methodology : Counter-Storytelling as an Analytical Framework for Educational Research." In *Foundations of Critical Race Theory in Education*, 159–74. Routledge, 2023. https://doi.org/10.4324/b23210-18.

Crossley, Nick, and Gemma Edwards. "Cases, Mechanisms and the Real: The Theory and Methodology of Mixed-Method Social Network Analysis." *Sociological Research Online*, May 31, 2016. https://doi.org/10.5153/sro.3920.

Csardi, Gabor, and Tamas Nepusz. "The Igraph Software Package for Complex Network Research."

*InterJournal* Complex Systems (2006): 1695.
Dou, Remy, and Justyna P. Zwolak. "Practitioner's Guide to Social Network Analysis: Examining Physics Anxiety in an Active-Learning Setting." *Physical Review Physics Education Research* 15, no. 2 (July 3, 2019): 020105. https://doi.org/10.1103/PhysRevPhysEducRes.15.020105.
Freeman, Elizabeth. *Time Binds: Queer Temporalities, Queer Histories*. Duke University Press, 2010. https://doi.org/10.1215/9780822393184.
Goertzen, Renee Michelle, Eric Brewe, and Laird Kramer. "Expanded Markers of Success in Introductory University Physics." *International Journal of Science Education* 35, no. 2 (2013): 262–88. https://doi.org/10.1080/09500693.2012.718099.
Gutzwa, Justin A., Ramón S. Barthelemy, Camila Amaral, Madison Swirtz, Adrienne Traxler, and Charles Henderson. "How Women and Lesbian, Gay, Bisexual, Transgender, and Queer Physics Doctoral Students Navigate Graduate Education: The Roles of Professional Environments and Social Networks." *Physical Review Physics Education Research* 20, no. 2 (September 12, 2024): 020115. https://doi.org/10.1103/PhysRevPhysEducRes.20.020115.
Hanneman, Robert A., and Mark Riddle. "Concepts and Measures for Basic Network Analysis." In *The SAGE Handbook of Social Network Analysis*, edited by John Scott and Peter J. Carrington. London ; Thousand Oaks, Calif: SAGE, 2011.
Hogan, Bernie, Juan Antonio Carrasco, and Barry Wellman. "Visualizing Personal Networks: Working with Participant-Aided Sociograms." *Field Methods* 19, no. 2 (May 1, 2007): 116–44. https://doi.org/10.1177/1525822X06298589.
Hollstein, Betina, Tom Töpfer, and Jürgen Pfeffer. "Collecting Egocentric Network Data with Visual Tools: A Comparative Study." *Network Science* 8, no. 2 (June 2020): 223–50. https://doi.org/10.1017/nws.2020.4.
Horkheimer, Max. *Critical Theory: Selected Essays*. New York: Continuum Pub. Corp, 1982.
Hyater-Adams, Simone, Claudia Fracchiolla, Noah Finkelstein, and Kathleen Hinko. "Critical Look at Physics Identity: An Operationalized Framework for Examining Race and Physics Identity." *Physical Review Physics Education Research* 14, no. 1 (June 1, 2018): 010132. https://doi.org/10.1103/PhysRevPhysEducRes.14.010132.
Krenz, Till. "Using Egor to Analyse Ego-Centered Network Data," February 1, 2024. https://cran.r-project.org/web/packages/egor/vignettes/using_egor.html.
Krenz, Till, Pavel N. Krivitsky, Raffaele Vacca, Michal Bojanowski, and Andreas Herz. "Egor: Import and Analyse Ego-Centered Network Data," July 2, 2018. https://doi.org/10.32614/CRAN.package.egor.
Lather, Patti. "Research as Praxis." *Harvard Educational Review* 56, no. 3 (September 1, 1986): 257–78. https://doi.org/10.17763/haer.56.3.bj2h231877069482.
Marin, Alexandra, and Barry Wellman. "Social Network Analysis: An Introduction." In *The SAGE Handbook of Social Network Analysis*, edited by John Scott and Peter J. Carrington. SAGE Publications, 2014. https://methods.sagepub.com/book/the-sage-handbook-of-social-network-analysis.
Marsden, Peter V. "Network Data and Measurement." *Annual Review of Sociology* 16 (1990): 435–63.
McCarty, Christopher, and Amber Wutich. "Conceptual and Empirical Arguments for Including or Excluding Ego from Structural Analyses of Personal Networks." *Connections* 26, no. 2 (2005): 82–88.
McCormick, Melinda, Ramon Barthelemy, and Charles Henderson. "Women's Persistence into Graduate Astronomy Programs: The Roles of Support, Interest, and Capital." *Journal of Women and Minorities in Science and Engineering* 20, no. 4 (2014). https://doi.org/10.1615/JWomenMinorScienEng.2014009829.
Olsen, Joe, Debbie Andres, Nicolette Maggiore, and Charles Ruggieri. "Characterizing Social Behavior Patterns in Teaching Assistant Interactions with Students." *Physical Review Physics Education Research* 19, no. 2 (September 13, 2023): 020129. https://doi.org/10.1103/PhysRevPhysEducRes.19.020129.

Pedersen, Thomas Lin. *Tidygraph: A Tidy API for Graph Manipulation*, 2024. https://tidygraph.data-imaginist.com.
Perry, Brea L., Bernice A. Pescosolido, and Stephen P. Borgatti. *Egocentric Network Analysis: Foundations, Methods, and Models*. 1st ed. Cambridge University Press, 2018. https://doi.org/10.1017/9781316443255.
Poel, Mart G. M. van der. "Delineating Personal Support Networks." *Social Networks* 15, no. 1 (March 1, 1993): 49–70. https://doi.org/10.1016/0378-8733(93)90021-C.
Pulgar, Javier, Diego Ramírez, Abigail Umanzor, Cristian Candia, and Iván Sánchez. "Long-Term Collaboration with Strong Friendship Ties Improves Academic Performance in Remote and Hybrid Teaching Modalities in High School Physics." *Physical Review Physics Education Research* 18, no. 1 (June 13, 2022): 010146. https://doi.org/10.1103/PhysRevPhysEducRes.18.010146.
"Qualitative Inquiry and Research Design | SAGE Publications Inc." Accessed May 13, 2024. https://us.sagepub.com/en-us/nam/qualitative-inquiry-and-research-design/book246896.
Quardokus, Kathleen, and Charles Henderson. "Promoting Instructional Change: Using Social Network Analysis to Understand the Informal Structure of Academic Departments." *Higher Education: The International Journal of Higher Education Research* 70, no. 3 (September 2015): 315–35. https://doi.org/10.1007/s10734-014-9831-0.
Quichocho, Xandria R., Erin M. Schipull, and Eleanor W. Close. "Understanding Physics Identity Development through the Identity Performances of Black, Indigenous, and Women of Color and LGBTQ+ Women in Physics," 412–17, 2020. https://www.per-central.org/items/detail.cfm?ID=15518.
R Core Team. *R: A Language and Environment for Statistical Computing*. Vienna, Austria: R Foundation for Statistical Computing, 2020. https://www.R-project.org/.
R Studio Team. "RStudio: Integrated Development for R." Boston, MA: PBC, 2020. http://www.rstudio.com/.
Rodriguez, Miguel, Ramón Barthelemy, and Melinda McCormick. "Critical Race and Feminist Standpoint Theories in Physics Education Research: A Historical Review and Potential Applications." *Physical Review Physics Education Research* 18, no. 1 (February 25, 2022): 013101. https://doi.org/10.1103/PhysRevPhysEducRes.18.013101.
Rosa, Katemari. "Educational Pathways of Black Women Physicists: Stories of Experiencing and Overcoming Obstacles in Life." *Physical Review Physics Education Research* 12, no. 2 (2016). https://doi.org/10.1103/PhysRevPhysEducRes.12.020113.
Rosa, Katemari, Jennifer Blue, Simone Hyater-Adams, Geraldine L. Cochran, and Chanda Prescod-Weinstein. "Resource Letter RP-1: Race and Physics." *American Journal of Physics* 89, no. 8 (August 1, 2021): 751–68. https://doi.org/10.1119/10.0005155.
Ryan, Louise, Jon Mulholland, and Agnes Agoston. "Talking Ties: Reflecting on Network Visualisation and Qualitative Interviewing." *Sociological Research Online* 19, no. 2 (May 1, 2014): 1–12. https://doi.org/10.5153/sro.3404.
Sundstrom, Meagan, David G. Wu, Cole Walsh, Ashley B. Heim, and N. G. Holmes. "Examining the Effects of Lab Instruction and Gender Composition on Intergroup Interaction Networks in Introductory Physics Labs." *Physical Review Physics Education Research* 18, no. 1 (January 4, 2022): 010102. https://doi.org/10.1103/PhysRevPhysEducRes.18.010102.
Swirtz, Madison, and Ramón Barthelemy. "Queering Methodologies in Physics Education Research." In *Physics Education Research Conference 2022*, 457–62. PER Conference. Grand Rapids, MI, 2022.
Traxler, Adrienne L., Camila Mani Dias Do Amaral, Charles Henderson, Evan LaForge, Chase Hatcher, Madison Swirtz, and Ramón Barthelemy. "Person-Centered and Qualitative Approaches to Network Analysis in Physics Education Research." *Physical Review Physics Education Research* 20, no. 2 (October 21, 2024): 020132. https://doi.org/10.1103/PhysRevPhysEducRes.20.020132.
Traxler, Adrienne L., Ximena C. Cid, Jennifer Blue, and Ramón Barthelemy. "Enriching Gender in PER:

A Binary Past and a Complex Future.” *Physical Review Physics Education Research* 12, no. 2 (August 1, 2016): 020114. https://doi.org/10.1103/PhysRevPhysEducRes.12.020114.
Wolf, Steven F., Timothy M. Sault, Tyme Suda, and Adrienne L. Traxler. “Social Network Development in Classrooms.” *Applied Network Science* 7, no. 1 (December 2022): 24. https://doi.org/10.1007/s41109-022-00465-z.
Wu, David G., Ashley B. Heim, Meagan Sundstrom, Cole Walsh, and N. G. Holmes. “Instructor Interactions in Traditional and Nontraditional Labs.” *Physical Review Physics Education Research* 18, no. 1 (March 14, 2022): 010121. https://doi.org/10.1103/PhysRevPhysEducRes.18.010121.

# Appendix A: Interview Protocol

1. Who is your current employer?
2. How long have you worked in your current position?
3. How did you end up in your current job?
4. How would you briefly describe your job?
5. What were your last 2–3 jobs, and roughly how long did you work at each of them?
6. Did you consider other career paths? [if yes:] Which ones? Why did you choose the path you took?
7. Suppose you need support with a major change in your career, such as changing jobs or pursuing a promotion. Who would you go to for advice about this change?
    A. Person A (B, C…) works at the same place as you?
    B. Approximately how long have you known them?
    C. How would you describe your relationship with A (B, C…)
    D. Do you connect with this person based on shared social identities or life experiences that are relevant to your relationship? And by shared social identities I mean any identities you and the people you name have in common, such as gender, race, ethnicity, profession, hobbies and so on [if no, do they have social identities which are different from yours and are relevant to your relationship?]

8. Suppose you need help with some part of your work, like covering tasks while you travel, or adding a new responsibility to your job. Who would you go to for advice or support in this?
    A. For how long have you known this person?

9 Do you socialize with people from your work? If the answer is “yes”: who do you socialize with?
    A. How would you describe your relationship with these people [say name by name]? For how long have you known them?

10. Are you part of any [formal or informal] identity-based advocacy or support groups? Examples might include a women in science group, professional organizations such as Out in STEM, or an allies group. [If yes:] Are any members of that group people who were listed in the above support and advice questions? [How would you describe your interactions with people from this/these groups? This group have regular activities or meetings? [If yes: can you tell me a little bit about these activities?]

11. Can you tell me about your key mentors in graduate school and how you met them?
12. Are there any people not already named who you would describe as current mentors, or people who helped you to find a job?
13. Can you tell me about what being a mentor means to you, and in case other names pop up, could you please talk about them?
14. Are there people outside of professional spaces who you would talk to about changing your career or to ask for career advice?
15. Are there people, either already on the diagram or not named yet, who you go to for emotional or moral support about job-related issues? [it can be someone inside or outside your work space]
16. [I can see in your answer on the survey that/as you mentioned before] you work in [sector the person works on]. Have you worked in other sectors before?
17. If the person had changed sectors before: Why did you change sectors? (Did you find what you were expecting in the sector you currently work in?)
18. How do you feel about the environment at your workplace?
19. Have you considered leaving your current career track? [If yes] Did you talk to anyone about it? Who did you talk to? How would you classify your relationship with these people? Did they give you some kind of advice about it? What was the advice? Did it influence your decision?
20. Would you recommend your field to others?